\documentclass[aps,prx,reprint,showpacs,superscriptaddress,floatfix]{revtex4-1}
\usepackage{comment}
\usepackage{graphics,amssymb,amsmath,epsfig,color,textgreek}
\usepackage{graphicx}
\usepackage[dvipsnames]{xcolor}
\usepackage{dcolumn}
\usepackage{bm}
\usepackage[colorlinks=true,citecolor=cyan]{hyperref}
\hypersetup{colorlinks=true,citecolor=cyan,linkcolor=red,urlcolor=magenta}
\usepackage{braket}
\usepackage[normalem]{ulem}
\usepackage{cancel}
\usepackage{diagbox}
\usepackage{lipsum}
\usepackage{algorithm}
\usepackage[noend]{algpseudocode}
\usepackage{ORCIDinREVTeX}
\usepackage{cleveref}
\usepackage{orcidlink}

\crefformat{figure}{Fig.~#2#1#3}
\crefformat{equation}{Eq.~#2#1#3}
\crefformat{appendix}{App.~#2#1#3}
\crefformat{section}{Sec.~#2#1#3}

\usepackage[dvipsnames]{xcolor}

\newcommand{\hc}{\text{H.c.}}

\newcommand{\edit}[1]{\textcolor{black}{#1}}

\newcommand{\fullbath}{\Lambda}

\newcommand{\bi}{\begin{itemize}}
\newcommand{\ei}{\end{itemize}}
\newcommand{\be}{\begin{equation}}
\newcommand{\ee}{\end{equation}}

\newcommand{\imag}{i}

\newcommand{\mcS}{\mathcal{S}}
\newcommand{\mcG}{\mathcal{G}}

\newcommand{\mcM}{\mathcal{M}}
\newcommand{\phant}{{\phantom{\dagger}}}
\newcommand{\mbf}[1]{\mathbf{#1}}

\begin{document}

\title{Efficient Quantum Implementation of Dynamical Mean Field Theory \\ for Correlated Materials}

\author{Norman~Hogan (corresponding author)\orcidlink{0000-0003-0720-4949}}
\email{anhogan3@ncsu.edu}
\affiliation{Department of Physics and Astronomy, North Carolina State University, Raleigh, North Carolina 27695, USA}

\author{Efekan~K\"okc\"u\orcidlink{0000-0002-7323-7274}}
\email{efekan.kokcu@ucf.edu}
\affiliation{Applied Mathematics and Computational Research Division,
            Lawrence Berkeley National Laboratory,
            Berkeley, CA 94720, USA}
\affiliation{Department of Electrical and Computer Engineering, University of Central Florida, Orlando, Florida 32816, USA}

\author{Thomas~Steckmann\orcidlink{0000-0001-6012-2948}}
\affiliation{Joint Center for Quantum Information and Computer Science,
University of Maryland and NIST, College Park, Maryland 20742, USA}

\author{Liam~P.~Doak\orcidlink{0009-0000-7326-6502}}
\affiliation{Department of Physics and Astronomy, North Carolina State University, Raleigh, North Carolina 27695, USA}

\author{Carlos~Mejuto-Zaera\orcidlink{0000-0001-5921-0959}}
\affiliation{Univ Toulouse, CNRS, Laboratoire de Physique Théorique, Toulouse, France}

\author{Daan~Camps\orcidlink{0000-0003-0236-4353}}
\affiliation{National Energy Research Scientific Computing Center,
            Lawrence Berkeley National Laboratory,
            Berkeley, CA 94720, USA}

\author{Roel~Van~Beeumen\orcidlink{0000-0003-2276-1153}}
\affiliation{Applied Mathematics and Computational Research Division,
            Lawrence Berkeley National Laboratory,
            Berkeley, CA 94720, USA}

\author{Wibe~A.~de~Jong\orcidlink{0000-0002-7114-8315}}
\affiliation{Applied Mathematics and Computational Research Division,
            Lawrence Berkeley National Laboratory,
            Berkeley, CA 94720, USA}

\author{A.~F.~Kemper\orcidlink{0000-0002-5426-5181}}
\email{akemper@ncsu.edu}
\affiliation{Department of Physics and Astronomy, North Carolina State University, Raleigh, North Carolina 27695, USA}

\date{\today}

\begin{abstract}

The accurate theoretical description of materials with strongly correlated electrons is a formidable challenge in condensed matter physics and computational chemistry. Dynamical Mean Field Theory (DMFT) is a successful approach that predicts behaviors of such systems by incorporating some of the correlated behavior using an impurity model, but it is limited by the need to calculate the impurity Green’s function. This work proposes a framework for DMFT calculations on quantum computers, focusing on near-term applications. It leverages the structure of the impurity problem, combining a low-rank Gaussian subspace representation of the ground state and a compressed, short-depth quantum circuit that joins state preparation with time evolution to compute Green’s functions. We demonstrate the convergence of the DMFT algorithm using the Gaussian subspace in a noise-free setting, and show the hardware viability of circuit compression by extracting the impurity Green’s function on IBM quantum processors for a single impurity coupled to three bath orbitals (8 qubits, 1 ancilla). We discuss potential paths toward realizing this quantum computing use case in materials science.
\end{abstract}

\maketitle

\section{Introduction}

\begin{figure*}
    \centering
    \includegraphics[width=0.96\linewidth]{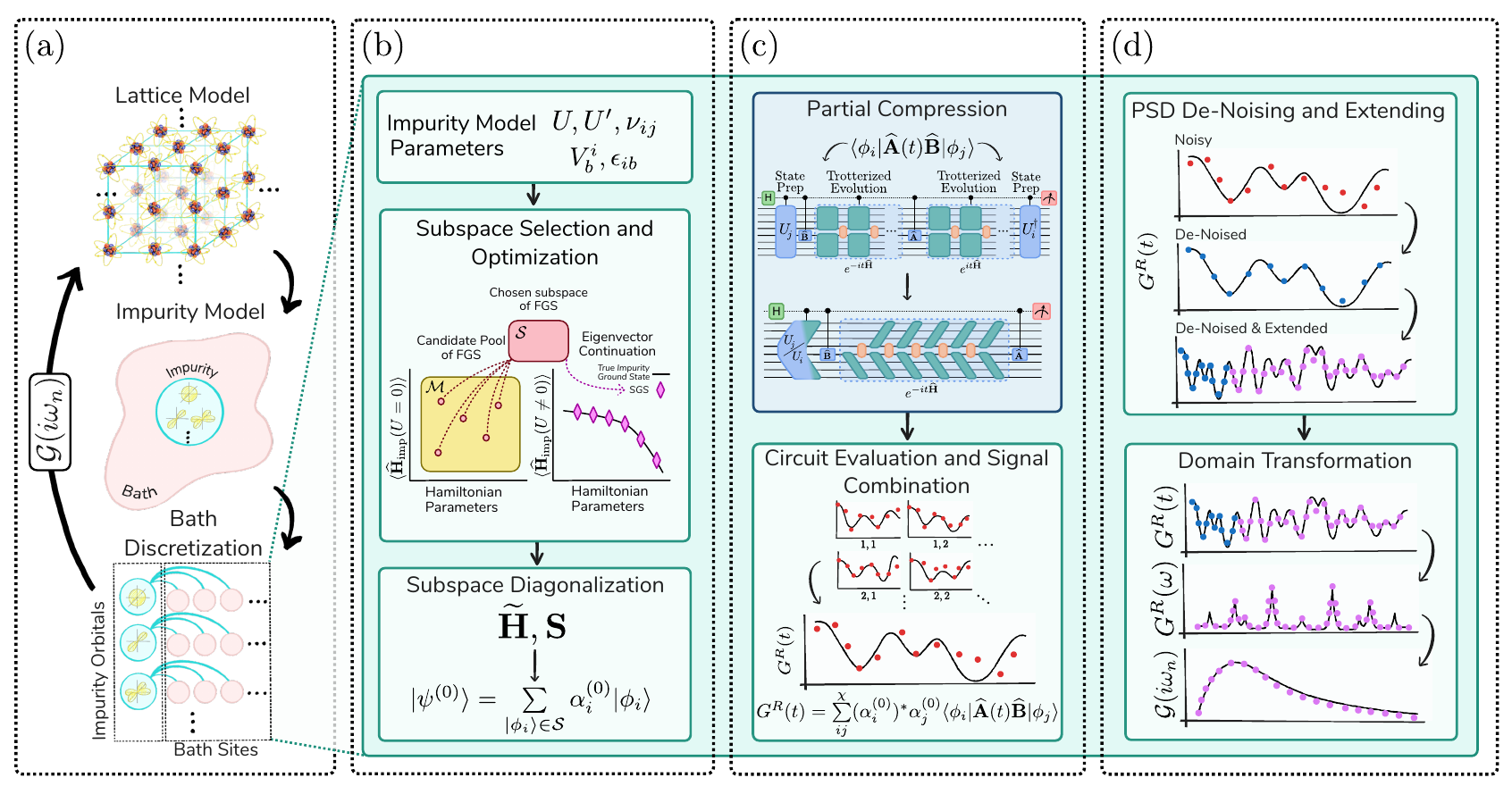}
    \caption{\textbf{Summary of this work.} (a) The DMFT embedding technique, which connects a complex lattice model of a real material to a simpler impurity model, requires the computation of the dynamics of the impurity model, often of the form of an impurity Green's function (GF), $\mathcal{G}(i\omega_n)$. (b) We approximate the impurity model's ground state with a sum of fermionic Gaussian states (FGS), which allows for efficient subspace diagonalization on classical hardware. This subspace, established at the start of the DMFT protocol, maintains high fidelity in subsequent iterations. (c) Time evolution with SGS is feasible on current quantum hardware via our partial compression technique, by exploiting the bath’s free-fermionic nature to reduce gate counts. Results are classically combined to recover a GF of the form $\langle \phi_i|\widehat{\mbf{A}}(t)\widehat{\mbf{B}}|\phi_j\rangle$. (d) After running circuits on noisy hardware with error mitigation, we use PSD de-noising and extending to extract key frequencies in the correlation function, providing the Matsubara GF $\mcG(i\omega_n)$ for the next DMFT iteration.}
    \label{fig:schematic}
\end{figure*}

Embedding methods~\cite{Georges1996DMFT,maier2005quantum, Chan_DMET, Zgid_SEET} are powerful frameworks that map many-body systems onto effective impurity problems coupled to a bath. In these approaches, a small cluster of strongly correlated electrons is embedded within a sea of mostly free electrons. Impurity models, such as the Anderson and Kondo impurity models familiar from condensed matter physics, provide the foundation for these embeddings. By adjusting the size of the correlated cluster, the system can be tuned continuously from nearly free to fully correlated, making these models computationally valuable. Strong correlations within the impurity are typically addressed using solvers like Quantum Monte Carlo (QMC)~\cite{Gull2011,Rubtsov2005}, Exact Diagonalization (ED)~\cite{Caffarel1994,capone2007solving,liebsch2011temperature}, and emerging quantum computing methods~\cite{keen2020quantum,steckmann2023mapping,Nie2024self-consistent,bauer2016hybrid,rungger2019dynamical,selisko2024dynamical,greenediniz2023quantum,besserve2022unraveling,jamet2022quantum, jamet2025anderson, ehrlichVQAbased}. The calculated information, often in the form of the single-particle Green’s function (GF), drives each iteration toward self-consistency between the fully correlated model and the impurity model.

Current impurity solvers used in embedding methods, including dynamical mean field theory (DMFT)~\cite{Georges1996DMFT,Kotliar2006,Paul2019,Kotliar2001a,Maier2005rev} as used here, face significant computational cost. QMC suffers from sign problems~\cite{Georges2013,Nomura2015}, ED faces exponential Hilbert space growth, and even tensor network and configuration interaction techniques only partially alleviate these limitations~\cite{Nunez2018,grundner_complex_2024,zgid_truncated_2012,go_adaptively_2017,mejuto-zaera_dynamical_2019}.
Quantum computing offers a potential path forward. Recent hybrid quantum-classical schemes for ground-state preparation, GF evaluation, and DMFT self-consistency demonstrate promising results on simulators and real devices~\cite{keen2020quantum,steckmann2023mapping,Nie2024self-consistent,bauer2016hybrid,rungger2019dynamical,selisko2024dynamical,greenediniz2023quantum,besserve2022unraveling,jamet2022quantum, jamet2025anderson}. However, most quantum methods are variational, affected by barren plateaus~\cite{mukherjee2023ComparativeStudyAdaptive,Ragone2024LieBarrenPlateaus} and sensitive to hardware noise. Time evolution typically demands deep circuits with thousands of two-qubit gates for moderate system sizes, challenging current hardware. A recent variational approach for time evolution of a single-impurity model~\cite{wolf2025variationaltimeevolutioncompression} reduces gate counts versus traditional Trotter evolution, but scaling analysis as one increases the size of the correlated cluster is needed.

Our approach develops a quantum-classical DMFT framework using low-depth circuits, exploiting efficient ground state representations as superpositions of Gaussian states (SGS)~\cite{bravyi2017complexity,boutin2021quantum}. This allows a classical determination of the ground state by subspace diagonalization and facilitates multiple DMFT iterations without changing the SGS basis (\cref{fig:schematic}(b)). We also employ algebraic circuit compression for both Gaussian state preparation and time evolution~\cite{kokcu2022algebraic,kokcu2024classification, camps2022algebraic} (\cref{fig:schematic}(c)). This strategy enables a noise-robust DMFT framework for both single impurity and multi-impurity models, and we demonstrate quantum hardware results of impurity GF calculations using error mitigation and physically motivated signal processing~\cite{kemper2024denoising} (\cref{fig:schematic}(d)), paving the way toward quantum advantage for correlated materials.

\section{Results}

\subsection{Subspace selection}\label{subsubsec: subspace selection}

\begin{figure}
    \centering
    \includegraphics[width=0.96\linewidth]{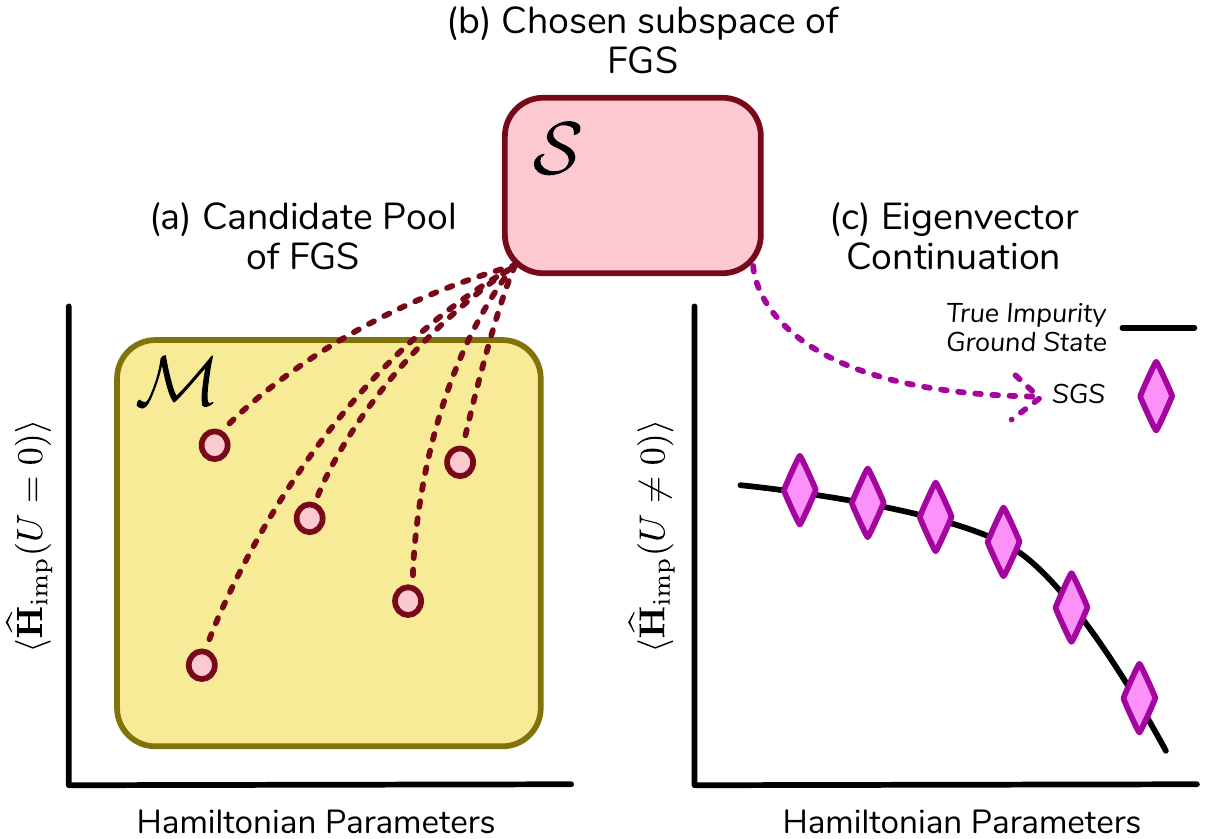}
    \caption{\textbf{Illustration of the subspace selection procedure.} (a) A candidate pool, $\mcM$, of FGS (shaded region) is generated, and from this, (b) a subspace $\mathcal{S}$ is chosen which approximates the ground state of some target impurity Hamiltonian $\mathbf{\widehat{H}}_\text{imp}$. (c) When solving for $\mathbf{\widehat{H}}_\text{imp}$ with nearby Hamiltonian parameters, the same subspace $\mathcal{S}$ is used to represent the impurity ground state using SGS, a technique called Eigenvector Continuation.}
    \label{fig: subspace algo}
\end{figure}

\begin{figure*}[ht]
    \centering
    \includegraphics[width=0.95\linewidth]{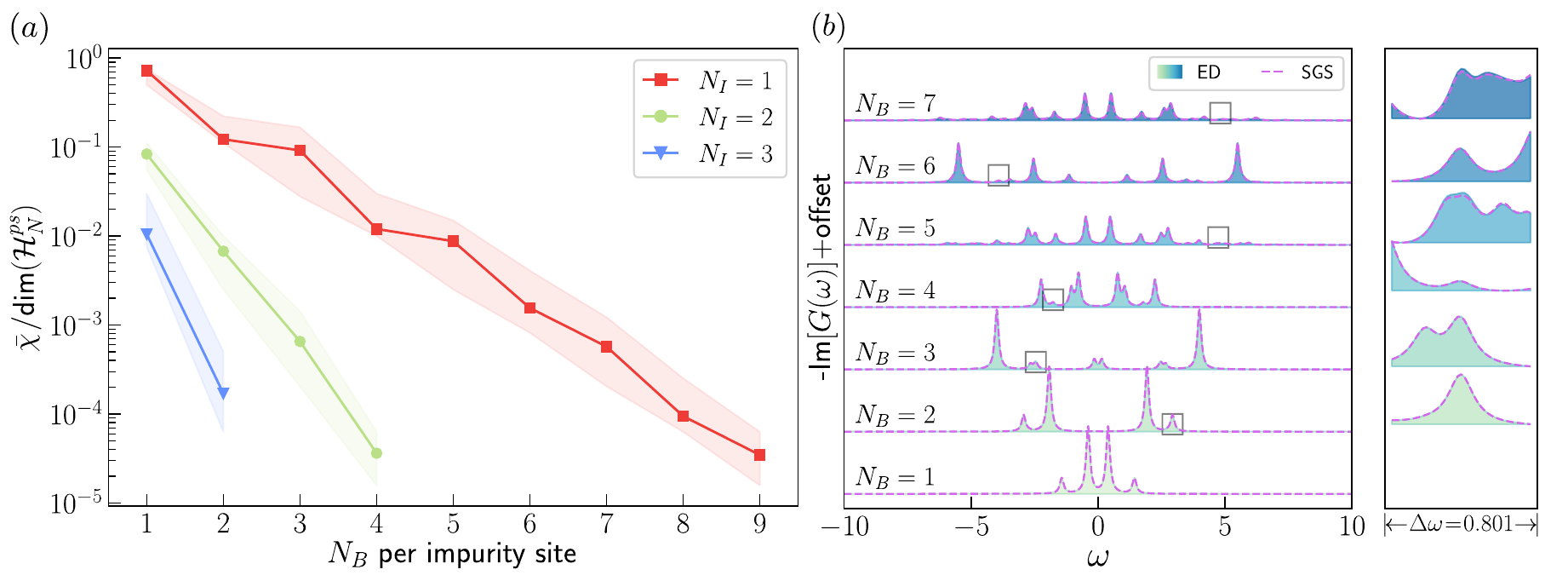}
    \caption{\textbf{Faithfullness of SGS as an approximation for the true impurity model ground state} (a) Fraction of the particle-selected Hilbert space $\mathcal{H}^{ps}_N$ needed for the SGS to reduce the relative error in the ground state energy to $\leq 10^{-2}$, averaged over 10 random sets of bath parameters \edit{and interaction strengths $U$}. The boundaries of the shaded region correspond to the \edit{minimum and maximum fraction across the runs}. (b, left) \edit{The impurity GF} for various system sizes computed with impurity ground states found through ED (shaded region) and those found using the SGS (dashed magenta line). (b, right) Zoomed-in view of small peaks in the full spectrum (regions on the left panel, boxed in grey).}
    \label{fig: SGS results}
\end{figure*}

We represent the impurity ground state as an SGS composed of a subspace of fermionic Gaussian states (FGS) for two main reasons: (i) they enable a classical, quasi-polynomial algorithm for approximating the ground state to arbitrary accuracy by adding FGS~\cite{bravyi2017complexity}, and (ii) they can be efficiently and exactly prepared on quantum hardware. Given many possible subspaces, one must choose an optimal set $\mcS$ that best approximates the ground state. The FGS in $\mcS$ should span the low-energy subspace of the impurity Hamiltonian $\widehat{\mbf{H}}_\text{imp}$ and be orthogonal enough to keep the overlap matrix $\mbf{S}$ invertible for subspace diagonalization. We identify such a subspace by incrementally adding FGS from a pre-generated \textit{candidate pool} $\mcM$ (\cref{fig: subspace algo}(a)), each chosen for orthogonality to previous selections (\cref{fig: subspace algo}(b)).

FGS in $\mcM$ are created by uniformly varying the non-interacting impurity problem parameters --- such as hopping strengths and on-site energies --- and setting the Coulomb repulsion parameter $U$ (or interaction strength) to zero. The first FGS, $\ket{\phi_1}$, minimizes $\langle \phi_1 | \widehat{\mbf{H}}_\text{imp} | \phi_1 \rangle$, and subsequent states maximize orthogonality to the subspace. Because the true impurity ground state is usually unknown, we check for convergence when the addition of further FGS yields negligible energy changes or all candidates in $\mcM$ are used or discarded due to ill-conditioning. We heuristically use a generous candidate pool of 1000 FGS, which ensures convergence for considered system sizes, though the optimal size is unknown. If the procedure stops due to ill-conditioning, increasing the pool size may be necessary.

Generating $\mcM$ is computationally cheap since each FGS can be represented as a $2N\times 2N$ covariance matrix for $N$ orbitals (where ``orbital" refers to a degree of freedom that allows two electrons to occupy), and computing expectation values and overlaps between non-orthogonal FGS is done by taking Pfaffians of, at most, $4\times4$ sub-matrices of the covariance matrix (see Appendix~\ref{asec: Gaussian states}). However, $\mcM$ may include highly non-orthogonal states, risking ill-conditioned eigenproblems. We check the condition number of $\mbf{S}$ and apply regularization methods~\cite{tikhonov1943stability} when needed.

The subspace $\mcS$ remains useful as impurity Hamiltonian parameters vary. According to techniques like Eigenvector Continuation~\cite{francis2022subspace, Mejuto_Zaera_2023, Frame_2018, yapa_volume_2022, drischler_toward_2021}, slowly changing parameters keep the ground state within a low-energy subspace, permitting reuse of $\mcM$ and $\mcS$ for different problem instances—such as in the DMFT protocol—see (\cref{fig: subspace algo}(c)). Significant parameter changes, however, may require re-selecting $\mcS$ from $\mcM$.

\subsection{Basis Convergence}\label{subsec: basis convergence}

We can now demonstrate the faithfulness of the SGS to the true impurity ground state found with ED. We assess convergence using two measures: how closely the SGS ground state energy ($\widetilde{E}^{(0)}$) approximates the true energy ($E^{(0)}$), and how well it reproduces features of the impurity GF ($G_\text{imp}^R$). The relative energy difference is

\begin{align}\label{eq: rel diff}
\mathcal{E}_{GS} = \left| \frac{\widetilde{E}^{(0)}-E^{(0)}}{E^{(0)}}\right|.
\end{align}

\begin{figure*}[ht]
    \centering
    \includegraphics[width=0.91\textwidth]{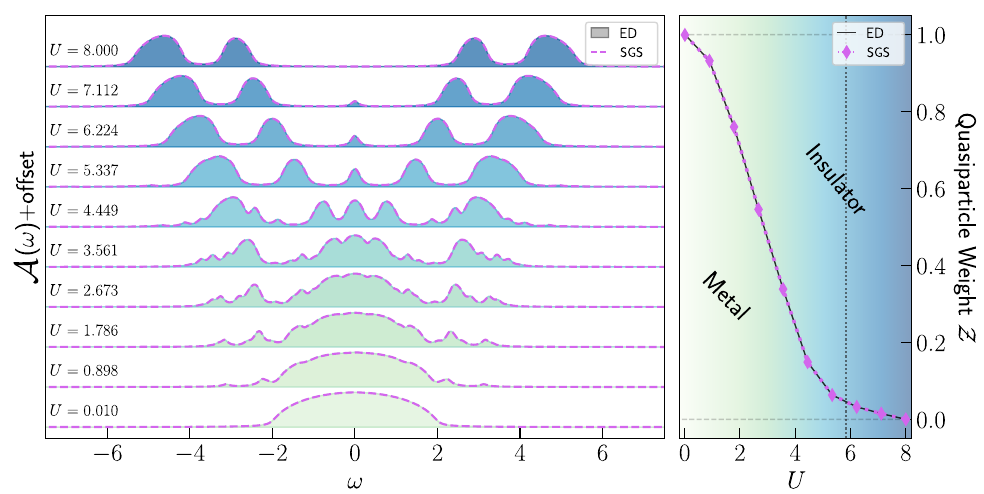}
    \caption{\textbf{Self-consistent results for a single-impurity model with $\mbf{N_B=3}$.} (left) The density of states (DOS) of the lattice model is computed using the self-consistently determined impurity self-energy found through DMFT using impurity ground states found with ED (shaded) and with SGS (magenta dashed line). (right) The self-consistent quasiparticle weight described by~\cref{eq: quasiparticle} for ED (black line) and the SGS (dashed magenta line with diamonds). \edit{The DMFT scan proceeds from the metallic to the insulating regime. The dashed vertical line indicates the continuum-bath critical value $U_c \approx 5.88$~\cite{Bulla1999dmftnrg}, providing a benchmark for the discrete ($N_B=3$) bath used here.}}
    \label{fig: lattice GFs and QP}
\end{figure*}

In~\cref{fig: SGS results}(a), we set $\mathcal{E}_{GS}\leq10^{-2}$ by truncating subspace selection, which determines the required SGS rank $\chi$ for this level of accuracy.
For our analysis, we consider multi-impurity models with a star topology: each of $N_I$ impurity orbitals connects to $N_B$ bath orbitals (see~\cref{fig: impurity hamiltonian}), yielding $N=N_I(N_B+1)$ total orbitals. We \edit{randomly sample interaction strengths $U$ between $0$ and $9$} and initialize 10 models with random bath parameters \edit{that respect particle-hole symmetry}. We restrict our analysis to impurity models at half-filling, and we find the average required rank $\overline\chi$ for a given system size $(N_I, N_B)$.

\cref{fig: SGS results}(a) shows $\overline\chi$ does not scale combinatorially with the particle-selected Hilbert space dimension $\text{dim}(\mathcal{H}^{ps}_N) = \binom{N}{ N/2}^2$ \edit{for even $N$ at half filling and $\text{dim}(\mathcal{H}^{ps}_N) = \binom{N}{ (N+1)/2}\binom{N}{(N-1)/2}$ for odd N}. Even as $\mathcal{H}^{ps}_N$ approaches the ED limit, $\overline\chi$ remains well below a percent, even with multiple impurity orbitals. \edit{The range of $\chi$ for~\cref{fig: SGS results}(a) is given explicitly in~\cref{ap: tables}}. In~\cref{fig: SGS results}(b), we let the subspace search run until termination \edit{(no bounds on $\mathcal{E}_{GS}$)} and calculate the impurity GF $G^R_\text{imp}(\omega)$ for single impurity models ($N_I=1$) with various bath sizes using ED \edit{and the Lehmann representation}. \edit{For odd N, we recover spin symmetry by averaging the spin-up and spin down impurity GF.} This computation requires the expansion of the SGS basis into the full Hilbert space. While this expansion is practical only for small systems, it demonstrates that convergence in energy leads to convergence in the state vector. The SGS closely matches ED results for $G^R_\text{imp}(\omega)$ in all cases, even with a few FGS (see~\Cref{tab: frac of hilbert space for impurity GFs}).

\begin{table}[ht]
    \centering
    \begin{tabular}{c|c|c|c}
         $N_B$ & $\text{dim}(\mathcal{H}^{ps}_N)$ & $\chi$ & $\chi/\text{dim}(\mathcal{H}^{ps}_N)$\\
         \hline
         1 & 4 & 3 & 0.75 \\
         2 & 9 & 6 & $0.67$ \\
         3 & 36 & 8 & $0.22$ \\
         4 & 100 & 16 & $0.16$ \\
         5 & 400 & 11 & $2.75\times 10^{-2}$ \\
         6 & 1225 & 16 & $1.31\times 10^{-2}$ \\
         7 & 4900 & 20 & $4.10\times 10^{-3}$ \\
    \end{tabular}
    \caption{\textbf{Fraction of Hilbert space required.} For the data in~\cref{fig: SGS results}(b), the rank of the SGS becomes a very small fraction of the particle-selected Hilbert space as the system size increases.}
    \label{tab: frac of hilbert space for impurity GFs}
\end{table}

Furthermore, both prominent and smaller high-frequency peaks are accurately reproduced by the SGS, as shown in the zoomed-in panel of~\cref{fig: SGS results}(b). Even for $N_B=7$, where the Hilbert space dimension far exceeds the SGS rank, $G^R_\text{imp}(\omega)$ remains accurate.

\subsection{DMFT convergence with SGS}\label{subsec: SGS DMFT}

The DMFT procedure requires an accurate, consistent representation of the impurity model’s dynamics as bath parameters vary from iteration to iteration. Before considering a quantum implementation of DMFT with SGS, we verify this accuracy in a noise-free setting.
We construct the optimal SGS subspace $\mcS$ using an initial $U$ and bath parameters and keep $\mcS$ fixed throughout the DMFT loop, so subspace optimization occurs only at initialization. This increases the importance of choosing good initial bath parameters. Heuristics --- such as monitoring DMFT convergence or periodic updates to $\mcS$ --- can help ensure $\mcS$ remains valid over parameter changes.

To confirm the validity of the initialized subspace, we performed self-consistent DMFT on the infinite-dimensional Hubbard model on a Bethe lattice with $N_I=1$ and $N_B=3$, a moderate problem size enabling ED comparison. \edit{The rank of each of the SGS across all $U$ upon DMFT convergence is given explicitly in~\cref{ap: tables}.} Across interaction regimes, the SGS used up to 24 FGS for $U=4.449$, representing a fairly significant particle-selected Hilbert space fraction ($\chi_C/\text{dim}(\mathcal{H}^{ps}_N)=0.67$). However, as shown in~\cref{fig: SGS results}(a), this fraction decreases as bath size increases. The DMFT-converged density of states (DOS) of the lattice, given by $\mathcal{A}(\omega) = -\frac{1}{\pi}\text{Im}[G_\text{latt}(\omega)]$, computed with SGS and ED ground states agree closely (see~\cref{fig: lattice GFs and QP} (left)), confirming that the SGS subspace reliably represents the impurity ground state throughout DMFT.

Following standard DMFT practice, in~\cref{fig: lattice GFs and QP} (right), we compare the quasiparticle weight $\mathcal{Z}$ --- an order parameter indicating the metal-to-insulator phase transition --- calculated from both the ED and SGS ground states. Because DMFT convergence is performed on the Matsubara frequency axis, we determine $\mathcal{Z}$ using

\begin{align}\label{eq: quasiparticle} 
    \mathcal{Z}^{-1} =  1 - \frac{\text{Im}[\Sigma_\text{imp}(i\omega_n)]}{\omega_n}\Bigg\vert_{\omega_n\rightarrow0}. 
\end{align}
Here, $\Sigma_\text{imp}(i\omega_n)$ denotes the self-energy of the impurity model, which represents the effect of correlations on its dynamics. DMFT seeks to ensure this self-energy matches the local self-energy of the fully-correlated lattice model. As with the lattice DOS $\mathcal{A}(\omega)$ shown in~\cref{fig: lattice GFs and QP} (left), the SGS and the exact impurity ground state yield consistent results from the DMFT procedure, demonstrating the SGS’s accuracy in capturing the impurity self energy and, by extension in the DMFT approximation, that of the lattice model.

\subsection{Partial Compression of Time Evolution Circuits for Impurity Models}
\label{sec:circuits}

\begin{figure}[ht!]
    \includegraphics[width=\columnwidth]{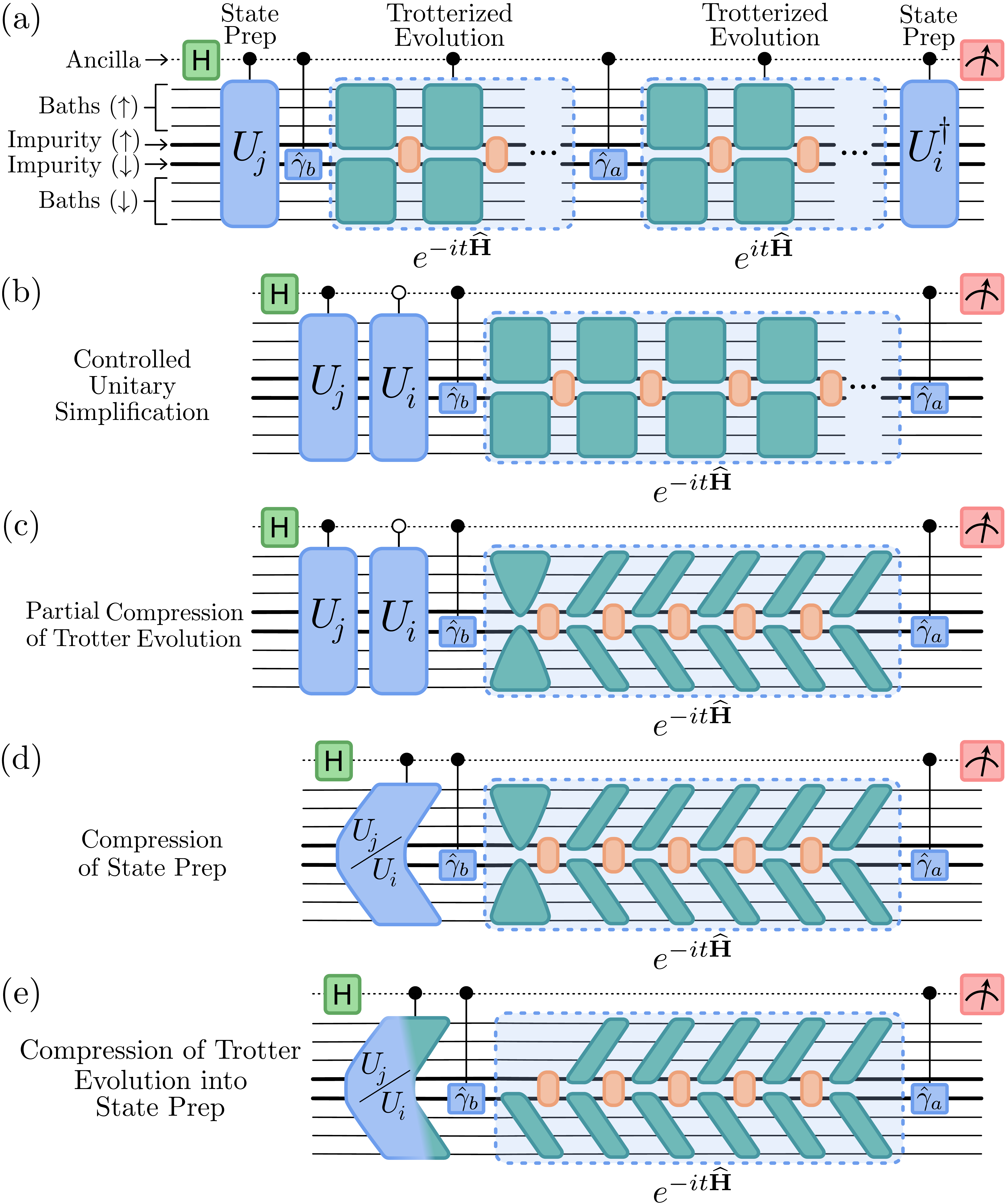}
    \caption{\textbf{Summary of partial compression.} (a) Typically, the circuit structure of state preparation and time evolution via Trotter decomposition is done like the topmost circuit. Our compression techniques involve first (b) simplifying the controlled unitary to bring the state preparation onto the leftmost side and (c) partially compressing the Trotter evolution. (d) Finally, the state prep of $\ket{\phi_i},\ket{\phi_j}$ is compressed further, and (e) some of the Trotter evolution is absorbed into state preparation.}
    \label{fig: circuit compression summary}
\end{figure}

To calculate the impurity GF, we need to evaluate the following overlaps

\begin{align}\label{eq:overlap2} 
\mathcal{C}_{ij}(t) = \langle \phi_i | \{\hat{d}^\phant_{\sigma}(t),\hat{d}_{\sigma}^\dag\} |\phi_j \rangle,
\end{align}
where $\ket{\phi_i}$ and $\ket{\phi_j}$ are FGS and $\hat{d}_{\sigma}^{(\dag)}$ are particle annihilation (creation) operators that act on a spin $\sigma$ within an impurity orbital. 
While the impurity GF for a general $N_I$ may be calculated with our methods, we limit the discussion to $N_I=1$ for our hardware runs. We also drop the spin index $\sigma$, since the problem considered has spin symmetry.

Once each of the $\mathcal{C}_{ij}(t)$ are calculated, we can classically recover the impurity GF with
\begin{align}\label{eq: classical combination}
    G^R_\text{imp}(t) = \sum_{i,j}^\chi (\alpha^{(0)}_i)^*\alpha_j^{(0)} \mathcal{C}_{ij}(t).
\end{align}

\begin{figure*}[th]
    \includegraphics[width=1.9\columnwidth]{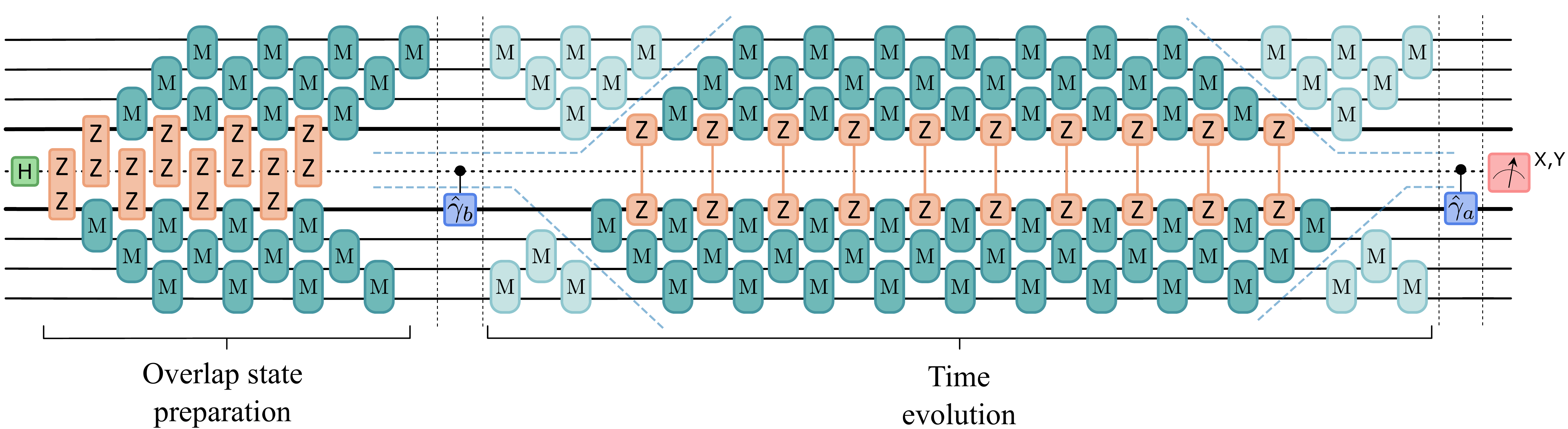}
    \caption{
    \textbf{Illustration of the circuit that produces the correlation $\mathbf{\bm{\langle \phi_i |}\hat{\gamma}_a(t) \hat{\gamma}_b\bm{|\phi_j\rangle}}$:}
    Here, $N_I = 1$, $N_B = 3$, and there are $r=10$ Trotter steps. The overlap state preparation and the time evolution parts are shown separately. The middle qubit is the ancilla qubit (dotted line), the impurity qubits are represented with bold lines, $H$ is a Hadamard gate, and the majority of the circuit consists of the matchgates (teal squares). The lighter-color matchgates at the beginning of the time evolution matchgates can pass through the controlled $\hat{\gamma}_b$ gate, and be absorbed by the state preparation matchgates. In addition, the lighter-color matchgates at the end of the circuit do not affect the measurement result, and therefore can be discarded. The necessary CNOTS for this simplified circuit acting on $N_q$ physical qubits (neglecting the ancilla) scales as $\mathcal{O}\left(N_q(N_q+rN_I)\right)$ in the $N_I \ll N_q$ limit.}
    \label{fig:overlap_circuits}
\end{figure*}

\noindent
where $\alpha^{(0)}_k$ are the ground state amplitudes for the FGS within the SGS, each of which are classically determined through subspace diagonalization.
Calculating $\mathcal{C}_{ij}(t)$ is typically accomplished using the Hadamard test, which we will use as well. The Hadamard test can be applied only for unitary operations; because the annihilation ($\hat{d}$) and creation ($\hat{d}^\dagger$) operators are not unitary, the overlap given in~\cref{eq:overlap2} cannot be directly calculated. For this purpose,
we recast the problem using unitary Majorana operators
($\hat{\gamma}_+=\hat{d}^\dag+\hat{d}^\phant$ and $\hat{\gamma}_-=i(\hat{d}\phant-\hat{d}^\dag)$ and calculate 
$\langle \phi_i | {\hat{\gamma}}_{a}(t){\hat{\gamma}}_{b} |\phi_j \rangle$ for $a,b \in \{ +, -\}$, which can be used to obtain $\mathcal{C}_{ij}(t)$ as
\begin{align}
    \mathcal{C}_{ij}(t) = \frac{1} {4}\sum_{\substack{a,b \in \{-,+\}}}
    s_{ab}\bra{\phi_i} \{ \hat{\gamma}_a(t), \hat{\gamma}_b \}  \ket{\phi_j},
\end{align}
where we defined $s_{++} = 1$, $s_{+-} = -i$, $s_{-+} = i$, $s_{--} = 1$. Each term in this summation can be obtained via a Hadamard test.

Typically, time evolution on quantum devices is done using Trotter product methods.
The schematic structure of the Hadamard test trotterized time evolution circuit for the evaluation of a correlation function
\begin{align}\label{eq: correlation function AB}
    \langle \phi_i |\hat{\gamma}_a(t) \hat{\gamma}_b|\phi_j\rangle
        = \langle 0|U_i e^{i \widehat{\mbf{H}} t} \hat{\gamma}_a e^{-i\widehat{\mbf{H}} t}\hat{\gamma}_bU^\dagger_j|0\rangle,
\end{align}
where $\hat{\gamma}_a(t)$ is the time evolved $\hat{\gamma}_a$ in the Heisenberg picture,
is shown in~\cref{fig: circuit compression summary}(a). We separate the system qubits into spin up and down subsets.

This is an efficient choice because the interactions (orange boxes) occur only on the impurity orbital(s), while the remaining operations (teal) are of the matchgate (equivalently, free-fermionic or Givens rotations) type. We define ``\textit{free fermionic matchgates}" or ``\textit{matchgates}" for short as the following two qubit gates

\begin{align}\label{eq: matchgate equation}
\begin{split}
    M_{i,i+1}(\vec{\theta}) =& e^{i \theta_1 Z_i} \: e^{i \theta_2 Z_{i+1}} \: e^{i \theta_3 X_i X_{i+1}} \\ &e^{i \theta_4 Y_i Y_{i+1}} \: e^{i \theta_5 Z_i } \: e^{i \theta_6 Z_{i+1}},
\end{split}
\end{align}
and is illustrated as
\begin{align}
    \centering
    \includegraphics[width = 0.67\columnwidth]{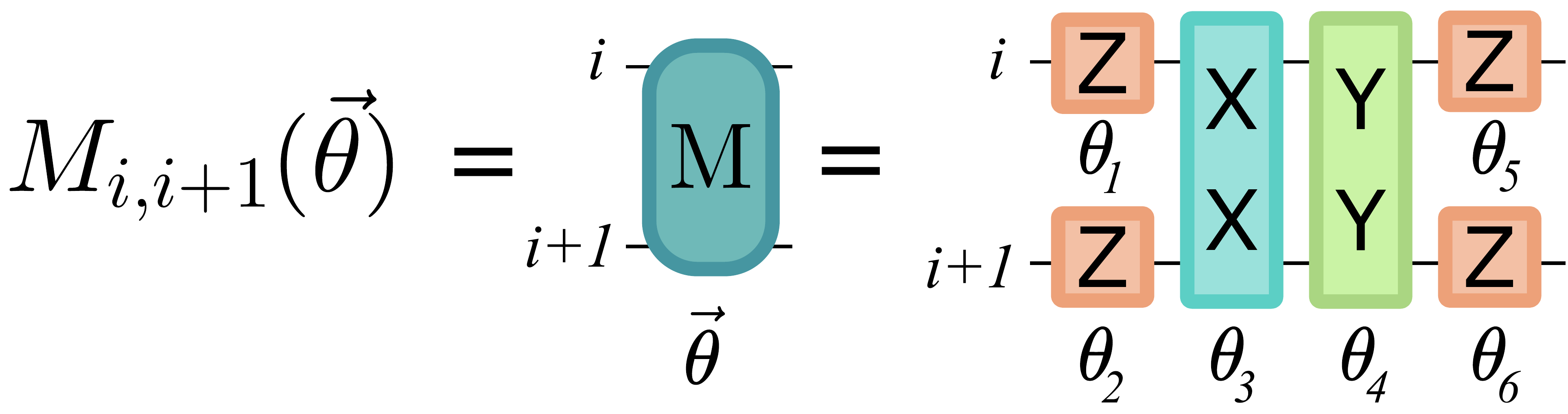}.
\end{align}

The state $\ket{\phi_i}$
is prepared via a controlled unitary operation, $U_i|0\rangle$. The first operator $(\hat{\gamma}_b)$ is then applied, controlled on the same ancilla. The system is time evolved to the desired time, followed by a similarly controlled second operator $(\hat{\gamma}_a)$ and conjugated preparation of the second state $\ket{\phi_j}$.
Now, because the states are Gaussian and the interactions are limited, we can significantly simplify and compress the circuit shown in ~\cref{fig: circuit compression summary}(a). We outline how this is achieved here --- a detailed discussion can be found in Appendix~\ref{app: partial compression details}.

Our first simplification leverages the control structure of the Hadamard test, focusing the measurement on the ancilla to (i) remove some controls, (ii) shift state preparation to the start of the circuit, and (iii) eliminate one time evolution operator, resulting in~\cref{fig: circuit compression summary}(b). For circuits with many time steps, these simplifications yield nearly a factor of 2 reduction in circuit depth. Each Trotter step can also be compressed into a compact (teal, triangle-shaped) structure of matchgate circuits~\cite{kokcu2022algebraic}. We will show that similar compression techniques can be applied to Trotter sections without interaction terms, enabling ``partial compression" that removes a significant number of gates, as illustrated in~\cref{fig: circuit compression summary}(c).

Because the states $\ket{\phi_i}$ and $\ket{\phi_j}$ are FGS, the controlled unitaries $U_i$ and $U_j$ that generate them can be written as controlled free fermionic evolution operators. These are then combined into a single controlled free fermionic evolution unitary with short depth~\cite{kokcu2023algebraic}, leading to~\cref{fig: circuit compression summary}(d).  Finally, we can absorb half of the initial Trotter step into the state preparation, leading to our final circuit structure in~\cref{fig: circuit compression summary}(e). The structure of the circuit that we ran on hardware is given by~\cref{fig:overlap_circuits} for $r=10$ trotter steps.


\subsection{Hardware Runs}\label{sec: hardware runs}

\begin{figure*}[th]
    \centering
    \includegraphics[width=0.92\textwidth]{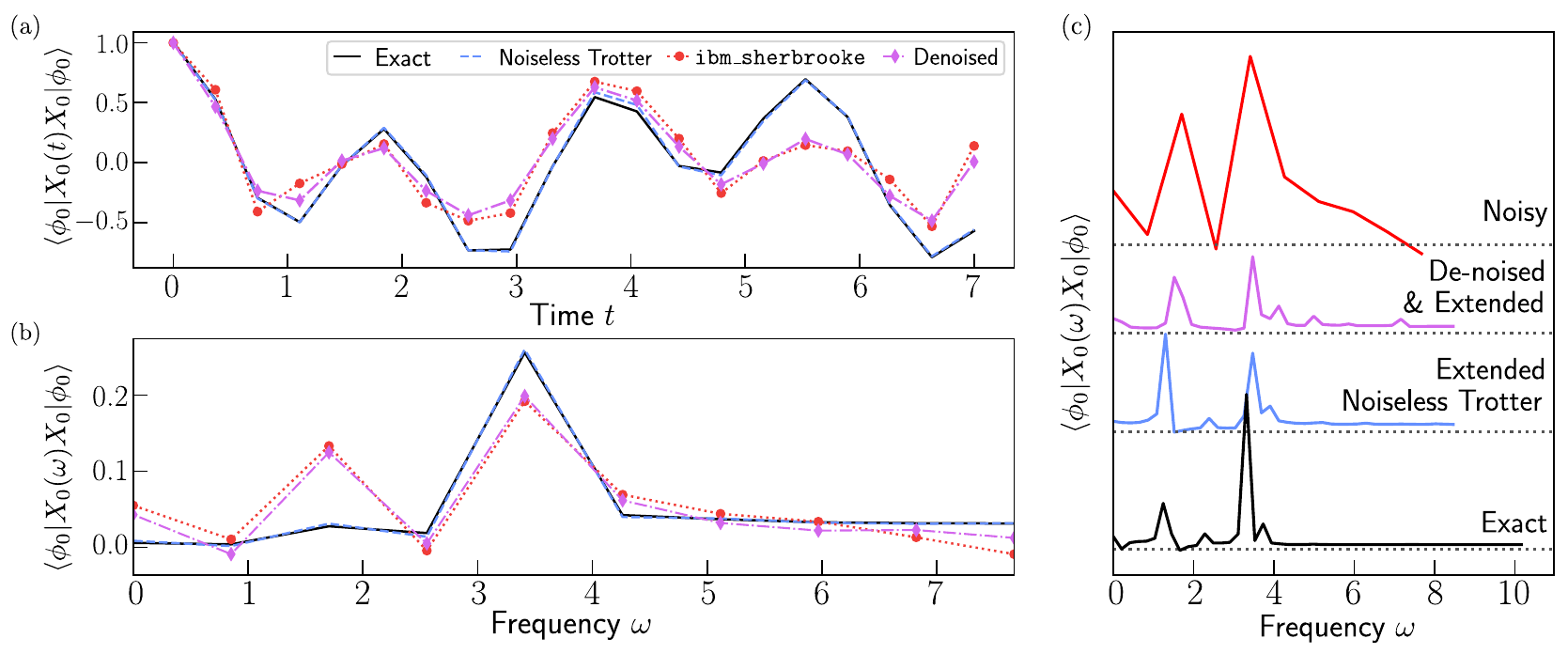}
    \caption{\textbf{Correlator computed on noisy quantum hardware.} Error-mitigated hardware results for the insulating phase ($U=5.337$) processed using PSD de-noising and extension for the $\ket{\phi_0}$-element of the real part of the $XX$-correlator on the impurity orbital from \texttt{ibm\_sherbrooke} shown in the (a) time domain and the (b)-(c) frequency domain. (a) The hardware data after error mitigation (red circles and dashed line). The solid black line is the evaluation of $G^R_\text{imp}(t)$ w.r.t. $\ket{\phi_0}$ using ED (labeled ``Exact"), the dashed blue line is noiseless circuit evaluations, and the diamond markers correspond to the hardware data after PSD de-noising. (b) Comparisons between the Fourier transforms of the data that are shown in (a). (c) The PSD-extended and de-noised hardware data (magenta line) compared to the error-mitigated hardware data without PSD de-noising (red line). The PSD-extended noiseless Trotter data (blue line) and the Fourier transform of the exact $G^R_\text{imp}(t)$ w.r.t. $\ket{\phi_0}$ evaluated for a longer time period (black line) are also shown for comparison.}
    \label{fig: sherbrooke mosaic}
\end{figure*}
Despite the significant reduction in the depth of our circuits that partial compression provides, various errors due to noisy hardware are anticipated. Although error mitigation was incorporated into our hardware runs, it is necessary to classically post-process the signal --- corresponding to the impurity GF --- to extract the dynamical information.

As discussed by~\cite{kemper2024denoising}, correlation functions of the form of~\cref{eq:overlap2}, such as the single impurity GF ($G^R_\text{imp}(t)$), are positive definite functions.
Numerically, $G^R_\text{imp}(t)$ is represented by a collection of discrete points on the time axis. The discretized $G^R_\text{imp}(t)$ can be arranged into a positive semi-definite (PSD) Hermitian Toeplitz matrix, commonly referred to as the ``Gram" matrix. 

If the data describing $G^R_\text{imp}(t)$ contains noise, as is often the case with data obtained from quantum hardware, the Gram matrix is no longer PSD. However, the noisy Gram matrix can be projected into the nearest PSD matrix, resulting in a de-noised, positive definite $G^R_\text{imp}(t)$. Another convenient feature of positive definite functions is that positive definite extensions exist to the finite data of $G^R_\text{imp}(t)$. Armed with the PSD de-noising and extending tools introduced in~\cite{kemper2024denoising}, we can both filter noise from $G^R_\text{imp}(t)$ and extend our data in the time domain without the need for additional quantum resources.

To demonstrate the performance of partial compression paired with the SGS on real quantum hardware, we evaluated the correlation function (see~\cref{eq: correlation function AB}) for
a single FGS in the SGS basis for a $N_I=1, N_B=3$ impurity model ($N_q=8$ plus an ancilla) in the insulating phase ($U=5.337$). From the Jordan-Wigner transformation, the full $G^R_\text{imp}(t)$ for this system is simplified to an $XX$-correlator (see App. B in Ref.~\cite{steckmann2023mapping}), requiring $\chi^2 n_t$ total circuit evaluations, where $n_t$ is the number of time points. In this example, the rank of the SGS is  $\chi=4$, and $n_t=20$. For this representation of the ground state of the impurity model, we require $\chi^2 n_t=320$ circuits to compute $G^R_\text{imp}(t)$. Details of the depth (in terms of CNOTs) of our circuits are provided in Appendix~\ref{ap: quantum resource estimation}.

For error mitigation, we used 10 cycles of Pauli twirling (sometimes called randomized compiling)~\cite{hashim2021randomized}, dynamical decoupling using an $XY$ pulse sequence~\cite{ezzell2023dynamical}, and rescaling based on gate count and error rates. \edit{With 10 cycles of Pauli twirling, our total number of circuits ran for the hardware results in~\cref{fig: sherbrooke mosaic} was 200}. Details on the various error mitigation techniques are included in Appendix~\ref{ap: post-processing}.

~\cref{fig: sherbrooke mosaic}(a) shows the error-mitigated noisy hardware results from \texttt{ibm\_sherbrooke}, and we compare them to the exact evaluation of $G^R_\text{imp}(t)$ and the noiseless Trotter results. The hardware results match the noiseless Trotter results reasonably well, which is confirmed by examining the Fourier transform in panel (b), where the prominent frequencies are present in both the noiseless Trotter and noisy data. 

We refine the hardware results by performing PSD de-noising and extension. The de-noising process seems to have a limited effect, as can be seen in panel (a).  However, the Fourier transform of the PSD de-noised and extended signal shown in panel (c) demonstrates that the extension sharpens the main features and picks up on the presence of an additional small peak on the shoulder of the large peak around $\omega=3.5$. \edit{Due to hardware noise, errors appear in the time-domain signal, resulting in the spurious frequency content seen at $\omega > 5$ under the Fourier transform in panel (c); these artifacts lie outside the physical bandwidth set by the discretized bath.}

DMFT implementations typically evaluate the impurity GF on the Matsubara (imaginary) frequency axis, where functions are smooth and amendable to numerical fitting. Although our methods for measuring the impurity GF on a quantum computer produce a GF in the time domain, obtaining the Matsubara GF is straightforward: performing a Fourier transform on $G^R_\text{imp}(t)$ reveals frequencies and amplitudes ($f_r$, $A_r$) using standard classical optimization. PSD extension grants access to longer time dynamics without additional quantum resources, improving Fourier resolution. The extracted ($f_r$, $A_r$) parameters are then used in~\cref{eq: reconstruct Matsubara GF} for the classical DMFT loop.

\begin{align}\label{eq: reconstruct Matsubara GF}
    \mcG_{\text{imp}}(i\omega_n) = \sum_r \frac{A_r}{i\omega_n - f_r}
\end{align}

\section{Discussion}

In this work, we have developed an efficient methodology for solving impurity problems with quantum hardware. A crucial ingredient for embedding theories, such as DMFT, is the simulation of the dynamics of the impurity model. It requires the ground state, which we have shown can be found economically through the combination of subspace methods alongside a basis of FGS. Further, the Gaussianity of this basis allows it to be exactly implemented on quantum hardware. More importantly, we have shown that this representation of the ground state remains faithful when implemented in the DMFT protocol. 

Our partial circuit compression significantly reduces the gate cost of both state preparation and time evolution by utilizing the free-fermionic nature of the bath and SGS, allowing for the absorption of some gates used for both state preparation and time evolution. The methods in this work reduce both the classical and quantum cost of computing the impurity GF, inviting further exploration into hybrid classical-quantum frameworks for leveraging embedding theories used in quantum chemistry and materials research.

While we have aimed to maximize computational efficiency, our methods have some limitations. Generating a candidate pool of FGS is more efficient than ED in memory and compute time, but it may still involve unnecessary classical calculations due to highly non-orthogonal states, leading to redundancies and ill-conditioning. Additionally, the initial guess for the bath parameterization is crucial; a poor guess can significantly alter the needed subspace for accurately representing the impurity ground state, impacting convergence in DMFT. Implementing a surrogate optimization method, as described in~\cite{herbst2022surrogate}, could streamline this process.

Another limitation is the reliance on today's quantum hardware, which lacks universal error correction and requires strategies to manage noise. Continuous efforts are underway to improve these systems at both the hardware and algorithmic levels. Classical post-processing algorithms are crucial for extracting signals from noisy data. \edit{The spurious high-frequency peaks visible in panel (c) of~\cref{fig: sherbrooke mosaic} illustrate one such artifact. Their impact on the full impurity GF, once the linear combination across FGS matrix elements is taken, is difficult to predict a priori and may affect DMFT convergence. A natural mitigation strategy exploits the discretized nature of the problem: since the bath is finite for any quantum implementation, the spectrum has a bounded number of poles, and peaks lying outside the bath bandwidth can be discarded as nonphysical. Additional mitigation comes from improvements in hardware fidelity and from averaging across multiple runs or different devices to suppress hardware-specific artifacts.} Unlike state-of-the-art methods like QMC and ED, partial circuit compression is not limited by system size or exponential sampling costs, aside from the capabilities of current qubit platforms. It is important to weigh the trade-off between hardware noise and the limitations in scaling the impurity model when using embedding techniques.

We believe that using a quantum computer as an impurity solver in DMFT could lead to early quantum advantage, given the particular nature of the problem.
In particular, retarded interactions are a possibility because time evolution is carried out using Trotterized circuits. It is feasible to explicitly incorporate time dependence into the Hamiltonian. However, further research is necessary on how to construct an effective bath, a topic that others have started to explore~\cite{schiro2019quantumimpurity,Scarlatella2021DMFT}. 
Since partial compression results in a more economical two-qubit gate cost, simulating an impurity cluster for cluster-DMFT is also a promising opportunity for today's quantum hardware.  
Overall, this approach moves us closer to accurately modeling real materials and molecular systems, as our methods readily integrate into more complex embedding schemes.

\section{Methods}

\subsection{Impurity Model}

\begin{figure}
    \centering
    \includegraphics[width=0.75\columnwidth]{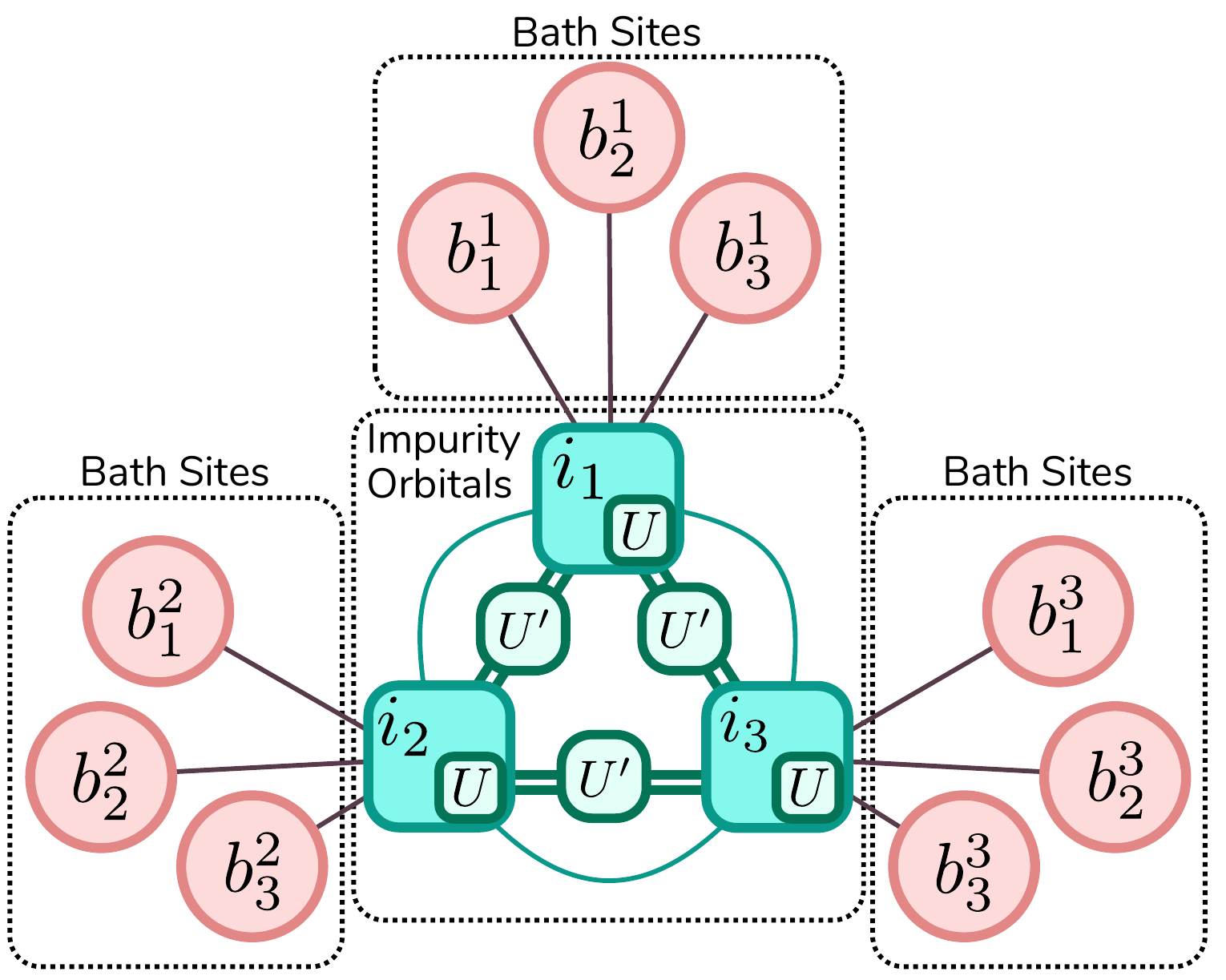}
    \caption{\textbf{Multiple impurity model Hamiltonian chosen for this work.} In this example, there are $N_I=3$ impurity orbitals (blue squares) and $N_B=3$ bath orbitals per impurity orbital (pink circles). Each orbital can support spin up and spin down. Single connecting lines denote inter-orbital hoppings, whereas double connecting lines between the impurity orbitals denote inter-impurity Coulomb interactions.}
    \label{fig: impurity hamiltonian}
\end{figure}

We begin by describing the multi-orbital impurity model,
with $N_I$ impurity orbitals and $N_B$ bath orbitals.
In this context, the term ``orbital" refers to a degree of freedom that allows two electrons to occupy.  In certain contexts, these could be literal orbitals,
for example, a $d-$orbital manifold; in others, these could be multiple
physical sites, such as used in cellular DMFT (cDMFT). The bath
topology is a second choice to be made.  Here, we use the star topology ---
each impurity orbital has an independent set of bath orbitals, each coupled only
to that impurity orbital (see~\cref{fig: impurity hamiltonian}).

The Hamiltonian that describes a multi-orbital impurity model has three components, $\widehat{\mbf{H}}_\text{imp} = \widehat{\mbf{H}}_{I}+\widehat{\mbf{H}}_B +  \widehat{\mbf{H}}_{IB}$.  The impurity-only terms are
\begin{subequations}\label{eq: impurity hamiltonian}
\begin{align}
\widehat{\mbf{H}}_{I} &=  
    \sum_{ij\sigma}^{N_I} \nu_{ij}\hat{d}^\dagger_{i\sigma}\hat{d}^\phant_{j\sigma} 
    +U \sum^{N_I}_{i}
    \hat{n}_{i\uparrow} \hat{n}_{i\downarrow} \nonumber \\
    & +U' \sum^{N_I}_{i\neq j}\sum_{\sigma\sigma'}
    \hat{n}_{i\sigma} \hat{n}_{j\sigma'}     
\end{align}

\noindent
where $\nu_{ij}$ are the on-site energies and intra-orbital hopping amplitudes, and we have distinguished the Coulomb interactions $U,U'$ to be intra- and inter-orbital, although this is not a limiting choice for our methods. The operators $\hat{d}_{i\sigma}^{(\dag)}$ are the fermionic annihilation (resp. creation) operators that act on a spin $\sigma \in \{\uparrow, \downarrow\}$ within an impurity orbital $i$. We have also used the number operator $\hat{n}_{i\sigma}=\hat{d}_{i\sigma}^{\dag}\hat{d}^\phant_{i\sigma}$.
The bath and the impurity-bath coupling terms are

\begin{align}
    \widehat{\mbf{H}}_B &= \sum_{i}^{N_I} \sum_{b\sigma}^{N_B} \epsilon_{ib} \hat{c}^\dagger_{ib\sigma}\hat{c}^\phant_{ib\sigma}  \\ 
    \widehat{\mbf{H}}_{IB} &= \sum_{i}^{N_I} \sum^{N_B}_{b\sigma}V^i_b (\hat{d}^\dagger_{i\sigma}\hat{c}^\phant_{ib\sigma} + \text{h.c.}),
\end{align}
\end{subequations}
Here, $\hat{c}_{ib\sigma}^{(\dag)}$ are the fermionic annihilation (creation) operators that act on a spin $\sigma$ within bath orbital $b$ that is connected to impurity orbital $i$. 

Note that only $\widehat{\mbf{H}}_I$ contains two-body terms, signifying that this model is non-interacting everywhere except for the impurity orbitals (as shown in~\cref{fig: impurity hamiltonian}). While $\widehat{\mbf{H}}_\text{imp}$ is presented here in its time-independent form, the inclusion of a dynamic Coulomb interaction $U(t)$, a requirement for incorporating dynamic screening effects, is readily compatible with our time evolution method.

The primary quantity we will focus on is the Green's function of the impurity, which is a key ingredient in DMFT as well as being directly related to photo-emission experiments. It is given by
\begin{align}\label{eq: impurity GF}
    [\mbf{G}^R_\text{imp}(t)]^\sigma_{i,j} &= -\imag \langle \Psi^{(0)} | \{\hat{d}^\phant_{i\sigma}(t),\hat{d}_{j\sigma}^\dag\} | \Psi^{(0)} \rangle
\end{align}
and it describes the single-particle excitations out of the ground
state~\cite{Mahan,bruus}.

Here, $\ket{\Psi^{(0)}}$ is the ground state of the impurity model with $i,j$ denoting impurity orbitals.
When the impurity model of interest contains more than one impurity orbital, the quantity $\mbf{G}^R_\text{imp}(t)$ is a $2N_I\times 2N_I$ matrix (where the factor of 2 comes from spin), each of which needs to be calculated.

\subsection{DMFT}

As discussed, a natural application of our methods for determining the ground state of the impurity model and evaluating the impurity GF through time evolution is in the context of DMFT. This embedding technique takes a strongly interacting lattice model, such as the Hubbard model, and self-consistently maps it to a local impurity model. The self-consistency is achieved when the impurity model self-energy matches the local self-energy of the original lattice model~\cite{Georges1996DMFT}.

In the limit of infinite lattice coordination for models with exclusively atomic interactions, the lattice self-energy $\Sigma_{\text{latt}}$ becomes local~\cite{Metzner1989,Mueller1989a,Mueller1989b,Georges1992}, i.e. independent of crystal momentum $\mbf{k}$, and in this limit the mapping to an impurity model is exact. Thus, for any finite-dimensional system, DMFT amounts to approximating the self-energy as a local quantity:

\begin{align}\label{eq: local self energy}
    \Sigma_{\text{latt}}(\omega,\mbf{k})\rightarrow \Sigma_{\text{latt,loc}}(\omega).
\end{align}

The self-consistency condition for DMFT then entails stating that the local lattice self-energy can be obtained from a suitably chosen impurity model. In other words, one can find an impurity model such that

\begin{align}\label{eq: self consistent}
    \Sigma_\text{imp} &\approx \Sigma_{\text{latt,loc}}
\end{align}
\noindent
where $\Sigma_{\text{latt,loc}}$ is the local self-energy of the fully interacting lattice model and $\Sigma_\text{imp}$ is obtained from Dyson's equation
\begin{align}\label{eq: dyson's eqn}
    \mbf{G} &= \mbf{G}^{0} + \mbf{G}^{0} \mbf{\Sigma} \mbf{G},
\end{align}

\noindent where $\mbf{G}^{0}$ is the non-interacting Green's function. 
At each iteration of DMFT, the current approximation of the lattice self-energy is evaluated using the assumption made with~\cref{eq: self consistent}. Throughout the DMFT loop, the parameters of the non-interacting bath of the impurity model are continuously updated until convergence on the level of the self-energy is reached. In Appendix~\ref{asec: DMFT details}, we provide a more in-depth overview of DMFT and the equations involved.

\subsection{Subspace Diagonalization}

Given an appropriate low-energy subspace, $\mcS=\{\ket{\phi_k}\}_{k=1}^\chi$, where $\ket{\phi_k}$ are FGS, the interacting ground state of the impurity model can be approximated with 

\begin{align}\label{eq: SGS}
    \ket{\psi^{(0)}} = \sum_{k=1}^{\chi}\alpha^{(0)}_k\ket{\phi_k}
\end{align}

\noindent
where $\chi$ is the rank (number of FGS) of the SGS. To find the coefficients $\alpha^{(0)}_k$, we solve the generalized eigenvalue problem in the subspace
%
\begin{align}\label{eq: subspace diagonalization}
    \widehat{\mbf{H}}\ket{\Psi^{(0)}}=E^{(0)}\ket{\Psi^{(0)}} &
    \quad \Rightarrow\quad \mbf{\widetilde{H}}\bm{\alpha}^{(0)}=\widetilde{E}^{(0)}\mbf{S}\bm{\alpha}^{(0)}
\end{align}
where $\ket{\Psi^{(0)}}$ is the ground state in the full Hilbert space, $\bm{\alpha}^{(0)}=[\alpha^{(0)}_1,...,\alpha^{(0)}_{\chi}]$ are the ground state coefficients in the SGS basis, and the matrix elements of 
$\mbf{\widetilde{H}}$ and $\mbf{S}$ are given by
$[\widetilde{\mbf{H}}]_{ij} = \langle \phi_i | \widehat{\mbf{H}} | \phi_j \rangle$, and $[\mbf{S}]_{ij} = \langle \phi_i | \phi_j \rangle$. 
Using the covariance matrix formalism described in Appendix ~\ref{asec: Gaussian states}, having a basis of FGS gives us a convenient and efficient way to evaluate these matrix elements classically.

\subsection{Partial compression}\label{subsec:Partical compression}

A major component of calculating a correlation function such as~\cref{eq:overlap2} is time evolution under the impurity Hamiltonian. 
To perform this time evolution efficiently, we make use of \textit{algebraic compression}, a recently developed circuit compression
algorithm~\cite{kokcu2022algebraic,camps2022algebraic,kokcu2023algebraic}. We outline the salient aspects here; a complete discussion can be found
in Appendix~\ref{app: partial compression details}. 

\begin{figure}[ht]
    \includegraphics[width=0.42\textwidth]{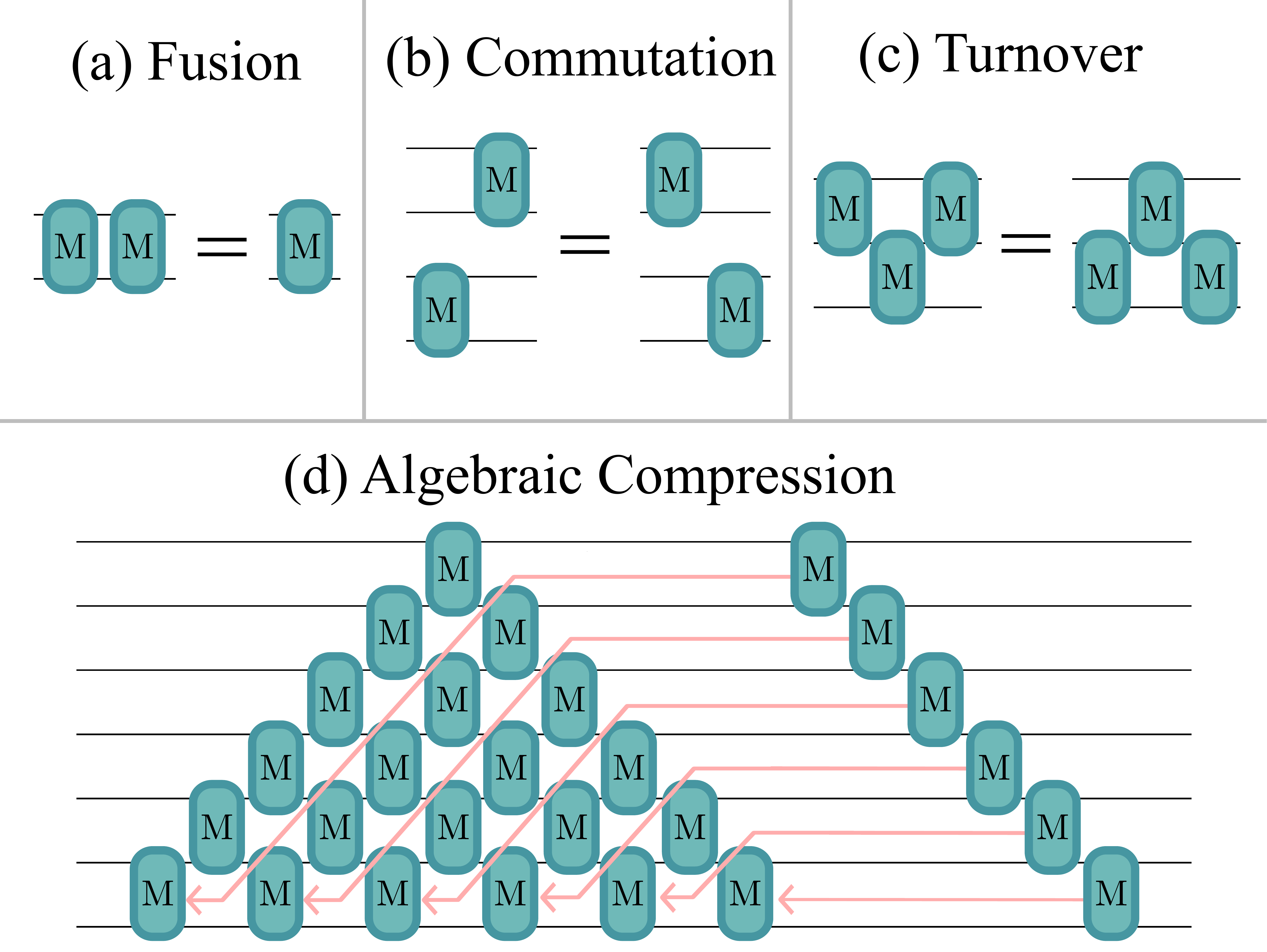}
    \caption{\textbf{Block properties that enable the algebraic compression of quantum circuits~\cite{kokcu2022algebraic,camps2022algebraic}.} 
    Matchgates satisfy the (a) fusion, (b) commutation and (c) turnover properties. Fusion and turnover properties involve a change in the angles of the gate, while commutation does not. These properties lead to (d) algebraic compression, which states that a triangle can absorb any matchgate. The arrows show how the incoming matchgates travel through the triangle.}
    \label{fig:rules}
\end{figure}


We will start by considering the definition of a matchgate given by~\cref{eq: matchgate equation}. The reason these are called ``free fermionic" is that after a Jordan-Wigner transformation, these gates correspond to a generic free fermionic interaction between adjacent qubits $i$ and $i+1$.

The matchgates satisfy three local algebraic properties: fusion, commutation, and turnover, as illustrated in~\cref{fig:rules}(a-c).
These properties are referred to as \textit{block rules}. The main theorem of~\cite{kokcu2022algebraic} states that any sequence of gates that satisfy the block rules can be compressed into a fixed-depth triangle structure via repeatedly applying these rules. An illustration of this for the matchgates is given in~\cref{fig:rules}(d). The arrows indicate how incoming matchgates traverse the triangle, during which the angles $\vec{\theta}$ are updated.
Matchgates are not the only set of gates satisfying the block properties. It was shown that the quantum gates used in the Trotter time evolution circuits of models such as 1-D TFIM, 1-D XY 
model~\cite{kokcu2022algebraic,camps2022algebraic},
as well as TFIM, TFXY, XY with periodic boundary conditions, and free fermions and controlled free fermions on any graph~\cite{kokcu2023algebraic} satisfy these or more extended versions of block properties. 

Algebraic compression can be applied to the Trotterized time evolution circuit of the following generic time dependent Hamiltonian:
\begin{align}
    \widehat{\mbf{H}}(t) = \widehat{\mbf{H}}_2(t) + \widehat{\mbf{H}}_4(t),
\end{align}

\noindent
where $\widehat{\mbf{H}}_2(t)$ consist of the quadratic terms, and $\widehat{\mbf{H}}_4(t)$ consist of the interaction terms given as follows:
\begin{align}\label{eq: H2 and H4}
\begin{split}
\widehat{\mbf{H}}_2(t) &= \sum_{\substack{i,j = 1 \\ \sigma \in \{\downarrow,\uparrow\}}}^{N_I + \fullbath} \big( h_{ij}(t) \, \hat{c}_{i \sigma}^\dagger \hat{c}_{j \sigma} + p_{ij}(t) \, \hat{c}_{i \sigma} \hat{c}_{j \sigma} + \hc\big), \\
\widehat{\mbf{H}}_4(t) &= \widehat{\mbf{H}}_4 \left( t ; \: \hat{d}_{1\sigma}, \hat{d}_{1\sigma}^\dagger, \hat{d}_{2\sigma}, \hat{d}_{2\sigma}^\dagger, \dots, \hat{d}_{N_I\sigma}, \hat{d}_{N_I\sigma}^\dagger \right) 
\end{split}
\end{align}
Here, $\fullbath$ denotes the total number of bath orbitals. In the case where there is an equal number of bath orbitals per impurity orbital (as in~\cref{eq: impurity hamiltonian}), then $\fullbath=N_IN_B$. We denote $\widehat{\mbf{H}}_4(t)$ to be a time dependent Hermitian operator which only has support over the impurity orbitals. It should be noted that since Eq. \eqref{eq: H2 and H4} is time dependent, it does not conserve the particle number due to the pair creation-annihilation terms in the quadratic part.

Let us consider the second-order Trotter time evolution circuit for this Hamiltonian, where one Trotter step is given as follows 
\begin{align}\label{eq: single trotter}
e^{-i\Delta t/2\: \widehat{\mbf{H}}_2(t)} e^{-i\Delta t \widehat{\mbf{H}}_4(t)} e^{-i\Delta t/2\: \widehat{\mbf{H}}_2(t)},
\end{align}
\noindent where $\Delta t = \frac{t}{r}$ and $r$ is the number of Trotter steps. It is shown that even though $\widehat{\mbf{H}}_2(t)$ contains long-range free fermionic interactions, its time evolution circuit
can be constructed via matchgates, and therefore
can be compressed to a triangle~\cite{kokcu2023algebraic}. This allows us to generate a circuit for $e^{-i\Delta t/2\: \widehat{\mbf{H}}_2(t)}$ with no approximation, yielding a Trotter error of order $\mathcal{O}(\mathrm{Poly}(N_I)t^3r^{-2})$ which is independent of the bath size. 

To determine the cost in terms of two-qubit (CNOT) gates, we will discuss 
the most general case, which also scales the worst: all impurity and bath qubits have nearest-neighbor hopping. We define $N_q$ as the total number of qubits, which for the impurity model is $N_q = 2(N_I + \fullbath)$.

\begin{figure*}[ht]
    \includegraphics[width=0.95\linewidth]{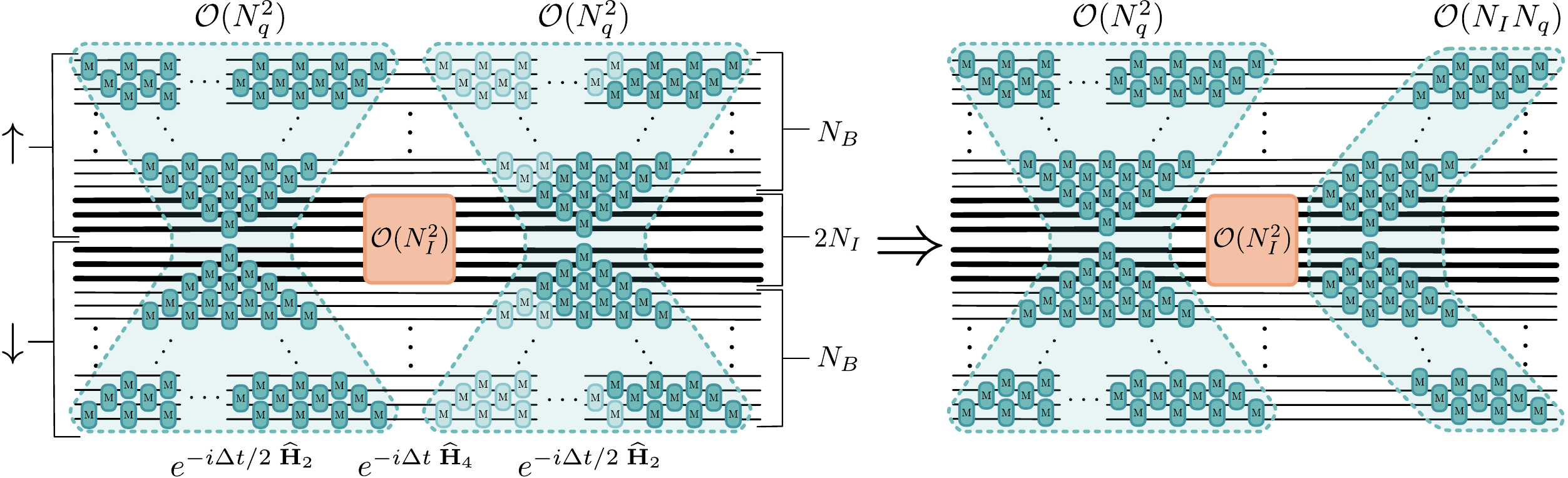}
    \caption{\textbf{Partial compression of a Trotter step.} Time evolution circuit structure for a single (left) uncompressed Trotter step with its equivalent (right) compressed form for $N_I=3$ and an arbitrary $N_B$. The labels $e^{-i\Delta t/2\phantom{.}\widehat{\mbf{H}}_2},e^{-i\Delta t/2\phantom{.}\widehat{\mbf{H}}_4}$ correspond to~\cref{eq: H2 and H4}. Matchgates are represented by the teal rectangles within the dotted lines. The lighter-colored matchgates signify those that are absorbed in partial compression, as shown by their absence in the circuit on the right. Adding additional Trotter steps results in another impurity block (labeled $\mathcal{O}(N_I^2)$) plus an $\mathcal{O}(N_IN_q)$ line of matchgates. The bold wires correspond to the impurity qubits, and the thinner lines belong to the bath qubits.}
    \label{fig: single trotter}
\end{figure*}

The circuit structure of the first Trotter step is shown on the left side of Fig. \ref{fig: single trotter}, where the number of CNOT gates is superimposed on the bath and impurity circuit blocks. Notice that a considerable amount---$\mathcal{O}(\fullbath^2)$--- of the matchgates commute with the impurity term, and therefore can be moved from the triangle on the right to the triangle on the left. Algebraic compression can then be applied to absorb these matchgates into the triangle on the left, which results in the circuit given on the right side of Fig.~\ref{fig: single trotter}.

In its uncompressed form, each Trotter step consists of four Triangles applied on $N_q$ (physical) qubits, leading to a total of $N_q\left(\frac{N_q}{2}-1\right)$ matchgates.
After applying the algebraic compression, it can be seen that the cost for one time step is reduced almost by half, and becomes $N_q\left(\frac{N_q}{2}-1\right)-\fullbath(\fullbath-1)$. 

A similar simplification occurs when we consider multiple time steps. In this case, the first time step can completely absorb an $\mathcal{O}(N_q^2)$ part of the second time step via algebraic compression, which reduces the cost of the second time step down to only $\frac{N_q}{2}\left(\frac{N_q}{2}-1\right)-\fullbath(\fullbath-1)$ matchgates. This cost per time step remains constant for any subsequent time step. 
The reduction is more prominent in the $N_I \ll N_q$ limit, as the resource requirement for additional Trotter steps is then reduced from $\mathcal{O}(N_q^2)$ to $\mathcal{O}(N_I N_q)$. A more detailed description of the resource estimation can be found in Appendix~\ref{ap: quantum resource estimation}.

The overlap state preparation circuit is obtained from a controlled free fermionic evolution. This leads to further simplifications when it is combined with the time evolution circuit. For an impurity Hamiltonian with $N_I = 1$ and $\fullbath=N_IN_B = 3$, we obtain the circuit given in Fig.~\ref{fig:overlap_circuits}, which has two sets of lighter-colored matchgates in the time evolution part of the circuit. The matchgates at the start of the time evolution can pass through the controlled $\hat{\gamma}_b$, and be absorbed by the triangle structures at the end of the overlap state preparation via algebraic compression. The other set of lighter-colored matchgates at the end of the circuit can be directly discarded simply because they do not affect the ancilla measurement result at the end. After these simplifications, for $N_I$ impurity orbitals, $\fullbath$ bath orbitals, and $r$ time steps, the total number of CNOT gates required is reduced to the following:
\begin{align}
\begin{split}
\frac{1}{2}N_q^2+3N_q-2+r\left(6N_I^2-4N_I+4N_I\fullbath\right)
\end{split}
\end{align}
which scales as $\mathcal{O}\left(N_q(N_q+rN_I)\right)$ in the $N_I \ll N_q$ limit.
For the impurity model with $N_I = 1$ and $\fullbath = N_I N_B = 3$ examined in this work, our circuits contained up to $r=18$ Trotter steps, resulting in a maximum of 306 CNOT gates.


\hfill

\section*{Author Contributions}
NH and LPD performed the analysis of classical state preparation using fermionic Gaussian states. 
EK, DC, and RVB developed the partial compression methods and its application to impurity models.
NH, AFK, and CMZ established the theoretical framework to apply the methods presented in this work to Dynamical Mean Field Theory. 
NH, EK, and TS contributed to the construction of the circuits and the analysis of data from the hardware runs. NH, EK, TS, and AFK were responsible for writing the manuscript. LPD, CMZ, DC, RVB, and WAdJ revised and provided feedback on the final manuscript.

\begin{acknowledgments}
We acknowledge helpful discussions with Steve Johnston.
NH and AFK were supported by the U.S. National Science Foundation under Grant No.~DMR-1752713.
EK, DC, and RVB were supported by the U.S. Department of Energy (DOE) under Contract No.~DE-AC02-05CH11231 through the Office of Advanced Scientific Computing Research  Accelerated Research for Quantum Computing  Program. WAdJ acknowledges support from the “Embedding Quantum Computing into Manybody Frameworks for Strongly Correlated Molecular and Materials Systems” project, by the U.S. DOE, Office of Science, Office of Basic Energy Sciences,  Division of Chemical Sciences, Geosciences, and Biosciences.  
\end{acknowledgments}

\section*{Data Availability}
The data that support the findings of this article are openly available~\cite{SGSDMFTdata}.

\section*{Declarations}

The authors declare no competing financial or non-financial interests.

\bibliography{CMZrefs,kemperlab}

@article{Kotliar2006,
	author={Gabriel Kotliar and Sergej Y. Savrasov and Kristjan Haule and Viktor S. Oudovenko and O. Parcollet and CA Marianetti},
	year={2006},
	title={Electronic structure calculations with dynamical mean-field theory},
	journal={Rev. Mod. Phys.},
	volume={78},
	number={3},
	pages={865},
        doi={10.1103/RevModPhys.78.865}
}

@article{Paul2019,
  title={Applications of dft+ dmft in materials science},
  author={Paul, Arpita and Birol, Turan},
  journal={Annual Review of Materials Research},
  volume={49},
  pages={31--52},
  year={2019},
  publisher={Annual Reviews},
  doi={10.1146/annurev-matsci-070218-121825}
}

@article{Maier2005rev,
	author={Thomas Maier and Mark Jarrell and Thomas Pruschke and Matthias H. Hettler},
	year={2005},
	title={Quantum cluster theories},
	journal={Rev. Mod. Phys.},
	volume={77},
	pages={1027},
        doi={10.1103/RevModPhys.77.1027}
}

@article{Kotliar2001a,
	author={Gabriel Kotliar and Sergej Y. Savrasov and Gunnar Palsson and Giulio Biroli},
	year={2001},
	title={Cellular dynamical mean field approach to strongly correlated systems},
	journal={Phys. Rev. Lett.},
	volume={87},
	number={18},
	pages={186401},
        doi={10.1103/PhysRevLett.87.186401}
}

@article{Rubtsov2005,
  title = {Continuous-time quantum Monte Carlo method for fermions},
  author = {Rubtsov, A. N. and Savkin, V. V. and Lichtenstein, A. I.},
  journal = {Phys. Rev. B},
  volume = {72},
  issue = {3},
  pages = {035122},
  numpages = {9},
  year = {2005},
  month = {Jul},
  publisher = {American Physical Society},
  doi = {10.1103/PhysRevB.72.035122},
  url = {https://link.aps.org/doi/10.1103/PhysRevB.72.035122}
}

@article{Georges2013,
  title={Strong correlations from Hund’s coupling},
  author={Georges, Antoine and Medici, Luca de' and Mravlje, Jernej},
  journal={Annu. Rev. Condens. Matter Phys.},
  volume={4},
  number={1},
  pages={137--178},
  year={2013},
  publisher={Annual Reviews},
  doi={10.1146/annurev-conmatphys-020911-125045}
}

@article{Nomura2015,
  title={Nonlocal correlations induced by Hund's coupling: A cluster DMFT study},
  author={Nomura, Yusuke and Sakai, Shiro and Arita, Ryotaro},
  journal={Physical Review B},
  volume={91},
  number={23},
  pages={235107},
  year={2015},
  publisher={APS},
  doi={10.1103/PhysRevB.91.235107}
}

@article{Caffarel1994,
	author={Michel Caffarel and Werner Krauth},
	year={1994},
	title={Exact diagonalization approach to correlated fermions in infinite dimensions: Mott transition and superconductivity},
	journal={Phys. Rev. Lett.},
	volume={72},
	number={10},
	pages={1545},
        doi={10.1103/PhysRevLett.72.1545}
}

@article{Nunez2018,
  title={Solving the multi-site and multi-orbital dynamical mean field theory using density matrix renormalization},
  author={N{\'u}{\~n}ez Fern{\'a}ndez, Yuriel and Hallberg, K},
  journal={Frontiers in Physics},
  volume={6},
  pages={13},
  year={2018},
  doi={10.3389/fphy.2018.00013},
  url={https://doi.org/10.3389/fphy.2018.00013}
}

@article{Metzner1989,
	author={Walter Metzner and Dieter Vollhardt},
	year={1989},
	title={Correlated lattice fermions in $d=\infty$ dimensions},
	journal={Phys. Rev. Lett.},
	volume={62},
	number={3},
	pages={324},
        doi={10.1103/PhysRevLett.62.324}
}

@article{Mueller1989a,
	author={E. M\"uller-Hartmann},
	year={1989},
	title={Correlated fermions on a lattice in high dimensions},
	journal={Z. Phys. B Con. Mat.},
	volume={74},
	number={4},
	pages={507-512},
        doi={10.1007/BF01311397}
}

@article{Mueller1989b,
	author={E. M\"uller-Hartmann},
	year={1989},
	title={The Hubbard model at high dimensions: some exact results and weak coupling theory},
	journal={Z. Phys. B Cond. Mat.},
	volume={76},
	number={2},
	pages={211-217},
        doi={10.1007/BF01312686}
}

@article{mejuto-zaera_dynamical_2019,
	title = {Dynamical mean field theory simulations with the adaptive sampling configuration interaction method},
	volume = {100},
	issn = {2469-9950, 2469-9969},
	url = {https://link.aps.org/doi/10.1103/PhysRevB.100.125165},
	doi = {10.1103/PhysRevB.100.125165},
	language = {en},
	number = {12},
	urldate = {2024-01-10},
	journal = {Physical Review B},
	author = {Mejuto-Zaera, Carlos and Tubman, Norm M. and Whaley, K. Birgitta},
	month = sep,
	year = {2019},
	pages = {125165},
}

@article{zgid_truncated_2012,
	title = {Truncated configuration interaction expansions as solvers for correlated quantum impurity models and dynamical mean-field theory},
	volume = {86},
	issn = {1098-0121, 1550-235X},
	url = {https://link.aps.org/doi/10.1103/PhysRevB.86.165128},
	doi = {10.1103/PhysRevB.86.165128},
	language = {en},
	number = {16},
	urldate = {2024-01-10},
	journal = {Phys. Rev. B},
	author = {Zgid, Dominika and Gull, Emanuel and Chan, Garnet Kin-Lic},
	month = oct,
	year = {2012},
	pages = {165128},
}

@article{go_adaptively_2017,
	title = {Adaptively truncated {Hilbert} space based impurity solver for dynamical mean-field theory},
	volume = {96},
	copyright = {https://link.aps.org/licenses/aps-default-license},
	issn = {2469-9950, 2469-9969},
	url = {https://link.aps.org/doi/10.1103/PhysRevB.96.085139},
	doi = {10.1103/PhysRevB.96.085139},
	language = {en},
	number = {8},
	urldate = {2025-03-31},
	journal = {Phys. Rev. B},
	author = {Go, Ara and Millis, Andrew J.},
	month = aug,
	year = {2017},
	pages = {085139},
}

@article{grundner_complex_2024,
	title = {Complex time evolution in tensor networks and time-dependent {Green}'s functions},
	volume = {109},
	url = {https://link.aps.org/doi/10.1103/PhysRevB.109.155124},
	doi = {10.1103/PhysRevB.109.155124},
	number = {15},
	urldate = {2025-03-18},
	journal = {Phys. Rev. B},
	author = {Grundner, M. and Westhoff, P. and Kugler, F. B. and Parcollet, O. and Schollwöck, U.},
	month = apr,
	year = {2024},
	note = {Publisher: American Physical Society},
	pages = {155124},
}

@misc{SGSDMFTdata,
    title = {SGS-DMFT v1},
    author = {Hogan, N. and Doak, L. P.},
    year = {2025},
    doi = {10.5281/zenodo.17237866},
    url = {https://github.com/anormanhogan/SGS-DMFT}
}

@article{wolf2025variationaltimeevolutioncompression,
      title={Variational Time Evolution Compression for Solving Impurity Models on Quantum Hardware}, 
      author={Stefan Wolf and Martin Eckstein and Michael J. Hartmann},
      year={2025},
      eprint={2508.10526},
      journal={arXiv preprint},
      url={https://arxiv.org/abs/2508.10526}, 
}

@article{ehrlichVQAbased,
    title={Variational quantum-algorithm based self-consistent calculations for the two-site DMFT model on noisy quantum computing hardware},
    author={Jannis Ehrlich and Daniel F Urban and Christian Elsässer},
    year={2025},
    journal={J. Phys.: Condens. Matter},
    volume={37},
    issue = {22},
    pages = {225901},
    url={https://iopscience.iop.org/article/10.1088/1361-648X/add4db},
    doi={10.1088/1361-648X/add4db}}

@article{Bulla1999dmftnrg,
  title = {Zero Temperature Metal-Insulator Transition in the Infinite-Dimensional Hubbard Model},
  author = {Bulla, R.},
  journal = {Phys. Rev. Lett.},
  volume = {83},
  issue = {1},
  pages = {136--139},
  numpages = {0},
  year = {1999},
  month = {Jul},
  publisher = {American Physical Society},
  doi = {10.1103/PhysRevLett.83.136},
  url = {https://link.aps.org/doi/10.1103/PhysRevLett.83.136}
}

@article{johnson1988mixing,
  title = {Modified Broyden's method for accelerating convergence in self-consistent calculations},
  author = {Johnson, D. D.},
  journal = {Phys. Rev. B},
  volume = {38},
  issue = {18},
  pages = {12807--12813},
  numpages = {0},
  year = {1988},
  month = {Dec},
  publisher = {American Physical Society},
  doi = {10.1103/PhysRevB.38.12807},
  url = {https://link.aps.org/doi/10.1103/PhysRevB.38.12807}
}

@article{Ragone2024LieBarrenPlateaus,
   title={A Lie algebraic theory of barren plateaus for deep parameterized quantum circuits},
   volume={15},
   ISSN={2041-1723},
   url={https://doi.org/10.1038/s41467-024-49909-3},
   DOI={10.1038/s41467-024-49909-3},
   number={1},
   journal={Nature Communications},
   publisher={Springer Science and Business Media LLC},
   author={Ragone, Michael and Bakalov, Bojko N. and Sauvage, Frédéric and Kemper, Alexander F. and Ortiz Marrero, Carlos and Larocca, Martín and Cerezo, M.},
   year={2024},
   month=aug }

@article{Georges1992,
  title = {Hubbard model in infinite dimensions},
  author = {Georges, Antoine and Kotliar, Gabriel},
  journal = {Phys. Rev. B},
  volume = {45},
  issue = {12},
  pages = {6479--6483},
  numpages = {0},
  year = {1992},
  month = {Mar},
  publisher = {American Physical Society},
  doi = {10.1103/PhysRevB.45.6479}
}

@article{steckmann2023mapping,
  title={Mapping the metal-insulator phase diagram by algebraically fast-forwarding dynamics on a cloud quantum computer},
  author={Steckmann, Thomas and Keen, Trevor and K{\"o}kc{\"u}, Efekan and Kemper, Alexander F and Dumitrescu, Eugene F and Wang, Yan},
  journal={Phys. Rev. Res.},
  volume={5},
  number={2},
  pages={023198},
  year={2023},
  publisher={APS},
  url={https://journals.aps.org/prresearch/abstract/10.1103/PhysRevResearch.5.023198}
}

@article{jamet2022quantum,
  title={Quantum subspace expansion algorithm for Green's functions},
  author={Jamet, Francois and Agarwal, Abhishek and Rungger, Ivan},
  journal={arXiv preprint},
  year={2022},
  url={https://arxiv.org/abs/2205.00094}
}

@book{Mahan,
    Author = {G. D. Mahan},
    Publisher = {Springer},
    Title = {Many Particle Physics},
    Address = {New York, NY10013, USA},
    Year = {2010},
    url = {https://doi.org/10.1007/978-1-4757-5714-9}
}

@book{bruus,
  title={Many-body quantum theory in condensed matter physics: an introduction},
  author={Bruus, Henrik and Flensberg, Karsten},
  year={2004},
  publisher={OUP Oxford},
  url={https://doi.org/10.1093/oso/9780198566335.001.0001}
}

@article{kokcu2022algebraic,
  title={Algebraic compression of quantum circuits for Hamiltonian evolution},
  author={K{\"o}kc{\"u}, Efekan and Camps, Daan and Bassman Oftelie, Lindsay and Freericks, James K and de Jong, Wibe A and Van Beeumen, Roel and Kemper, Alexander F},
  journal={Phys. Rev. A},
  volume={105},
  number={3},
  pages={032420},
  year={2022},
  publisher={APS},
  url={https://doi.org/10.1103/PhysRevA.105.032420}
}

@article{georges1996dynamical,
  title={Dynamical mean-field theory of strongly correlated fermion systems and the limit of infinite dimensions},
  author={Georges, Antoine and Kotliar, Gabriel and Krauth, Werner and Rozenberg, Marcelo J},
  journal={Reviews of Modern Physics},
  volume={68},
  number={1},
  pages={13},
  year={1996},
  publisher={APS}
}

@article{camps2022algebraic,
  title={An algebraic quantum circuit compression algorithm for hamiltonian simulation},
  author={Camps, Daan and K\"okc\"u, Efekan and Bassman Oftelie, Lindsay and De Jong, Wibe A and Kemper, Alexander F and Van Beeumen, Roel},
  journal={SIAM Journal on Matrix Analysis and Applications},
  volume={43},
  number={3},
  pages={1084--1108},
  year={2022},
  publisher={SIAM},
  url={https://doi.org/10.1137/21M1439298}
}

@article{Gull2011,
  title={Continuous-time quantum Monte Carlo impurity solvers},
  author={Emanuel Gull and Philipp Werner and Sebastian Fuchs and Brigitte Surer and Thomas Pruschke and
Matthias Troyer},
  journal={Computer Physics Communications},
  volume={182},
  pages={1078–1082},
  year={2011},
  url={https://journals.aps.org/rmp/abstract/10.1103/RevModPhys.83.349}
}

@article{maier2005quantum,
  title={Quantum cluster theories},
  author={Maier, Thomas and Jarrell, Mark and Pruschke, Thomas and Hettler, Matthias H},
  journal={Rev. of Mod. Phys.},
  volume={77},
  number={3},
  pages={1027},
  year={2005},
  publisher={APS},
  url={https://link.aps.org/doi/10.1103/RevModPhys.77.1027}
}

@article{Zgid_SEET,
  title={Generalized Self-Energy Embedding Theory},
  author={Tran Nguyen Lan and Dominika Zgid},
  journal={J. Phys. Chem. Lett.},
  volume={8},
  pages={2200-2205},
  year={2017},
  url={https://pubs.acs.org/doi/full/10.1021/acs.jpclett.7b00689}
}

@article{Chan_DMET,
  title={Density Matrix Embedding: A Simple Alternative to Dynamical Mean-Field Theory},
  author={Gerald Knizia and Garnet Kin-Lic Chan},
  journal={Phys. Rev. Lett.},
  volume={109},
  number={186404},
  year={2012},
  url={https://journals.aps.org/prl/abstract/10.1103/PhysRevLett.109.186404}
}

@article{bauer2016hybrid,
  author = {Bauer, Bela and Wecker, Dave and Millis, Andrew J. and Hastings, Matthew B. and Troyer, Matthias},
  doi = {10.1103/PhysRevX.6.031045},
  issue = {3},
  journal = {Phys. Rev. X},
  month = {Sep},
  pages = {031045},
  publisher = {American Physical Society},
  title = {Hybrid Quantum-Classical Approach to Correlated Materials},
  url = {https://link.aps.org/doi/10.1103/PhysRevX.6.031045},
  volume = {6},
  year = {2016},
}

@article{keen2020quantum,
  author = {Keen, Trevor and Maier, Thomas and Johnston, Steven and Lougovski, Pavel},
  doi = {10.1088/2058-9565/ab7d4c},
  journal = {Quantum Science and Technology},
  month = {Apr},
  number = {3},
  pages = {035001},
  publisher = {{IOP} Publishing},
  title = {Quantum-classical simulation of two-site dynamical mean-field theory on noisy quantum hardware},
  url = {https://doi.org/10.1088/2058-9565/ab7d4c},
  volume = {5},
  year = {2020},
}

@article{rungger2019dynamical,
  archiveprefix = {arXiv},
  author = {Rungger, I. and Fitzpatrick, N. and Chen, H. and Alderete, C. H. and Apel, H. and Cowtan, A. and Patterson, A. and Ramo, D. Munoz and Zhu, Y. and Nguyen, N. H. and Grant, E. and Chretien, S. and Wossnig, L. and Linke, N. M. and Duncan, R.},
  eprint = {1910.04735},
  primaryclass = {quant-ph},
  title = {Dynamical mean field theory algorithm and experiment on quantum computers},
  year = {2020},
  journal = {arXiv preprint},
  url = {https://arxiv.org/abs/1910.04735}
}

@article{francis2022subspace,
  title={Subspace Diagonalization on Quantum Computers using Eigenvector Continuation},
  author={Francis, Akhil and Agrawal, Anjali A and Howard, Jack H and K{\"o}kc{\"u}, Efekan and Kemper, Alexander F.},
  journal={arXiv preprint},
  year={2022},
  url={https://doi.org/10.48550/arXiv.2209.10571}
}

@article{Frame_2018,
  title={Eigenvector continuation with subspace learning},
  author={Frame, Dillon and He, Rongzheng and Ipsen, Ilse and Lee, Daniel and Lee, Dean and Rrapaj, Ermal},
  journal={Phys. Rev. Lett.},
  volume={121},
  number={3},
  pages={032501},
  year={2018},
  publisher={APS},
  url={https://doi.org/10.1103/PhysRevLett.121.032501}
}

@article{Mejuto_Zaera_2023,
  title={Quantum eigenvector continuation for chemistry applications},
  author={Mejuto-Zaera, Carlos and Kemper, Alexander F},
  journal={Electronic Structure},
  volume={5},
  number={4},
  pages={045007},
  year={2023},
  publisher={IOP Publishing},
  url={https://doi.org/10.1088/2516-1075/ad018f}

}

@article{yapa_volume_2022,
  title={Volume extrapolation via eigenvector continuation},
  author={Yapa, Nuwan and K{\"o}nig, Sebastian},
  journal={Phys. Rev. C},
  volume={106},
  number={1},
  pages={014309},
  year={2022},
  publisher={APS},
  url={https://doi.org/10.1103/PhysRevC.106.014309}
}

@article{drischler_toward_2021,
	title = {Toward emulating nuclear reactions using eigenvector continuation},
	volume = {823},
	issn = {03702693},
	url = {https://doi.org/10.1016/j.physletb.2021.136777},
	doi = {10.1016/j.physletb.2021.136777},
	urldate = {2022-04-10},
	journal = {Phys. Lett. B},
	author = {Drischler, C. and Quinonez, M. and Giuliani, P. G. and Lovell, A. E. and Nunes, F. M.},
	month = dec,
	year = {2021},
	pages = {136777},
}

@article{jamet2025anderson,
      title={Anderson impurity solver integrating tensor network methods with quantum computing}, 
      author={Francois Jamet and Connor Lenihan and Lachlan P. Lindoy and Abhishek Agarwal and Enrico Fontana and Baptiste Anselme Martin and Ivan Rungger},
      year={2025},
      journal={APL Quantum},
      volume={2},
      issue={1},
      doi={10.1063/5.0245488},
      url={https://doi.org/10.1063/5.0245488}
}

@article{greenediniz2023quantum,
  doi = {10.22331/q-2024-06-20-1383},
  url = {https://doi.org/10.22331/q-2024-06-20-1383},
  title = {Quantum {C}omputed {G}reen's {F}unctions using a {C}umulant {E}xpansion of the {L}anczos {M}ethod},
  author = {Greene-Diniz, Gabriel and Manrique, David Zsolt and Yamamoto, Kentaro and Plekhanov, Evgeny and Fitzpatrick, Nathan and Krompiec, Michal and Sakuma, Rei and Ramo, David Mu{\~{n}}oz},
  journal = {{Quantum}},
  issn = {2521-327X},
  publisher = {{Verein zur F{\"{o}}rderung des Open Access Publizierens in den Quantenwissenschaften}},
  volume = {8},
  pages = {1383},
  month = jun,
  year = {2024}
}

@article{boutin2021quantum,
  title = {Quantum impurity models using superpositions of fermionic Gaussian states: Practical methods and applications},
  author = {Boutin, Samuel and Bauer, Bela},
  journal = {Phys. Rev. Res.},
  volume = {3},
  issue = {3},
  pages = {033188},
  numpages = {15},
  year = {2021},
  month = {Aug},
  publisher = {American Physical Society},
  doi = {10.1103/PhysRevResearch.3.033188},
  url = {https://link.aps.org/doi/10.1103/PhysRevResearch.3.033188}
}

@article{bravyi2017complexity,
  title = {Complexity of {{Quantum Impurity Problems}}},
  author = {Bravyi, Sergey and Gosset, David},
  date = {2017-12-01},
  journal = {Communications in Mathematical Physics},
  volume = {356},
  number = {2},
  year = {2017},
  pages = {451--500},
  issn = {1432-0916},
  doi = {10.1007/s00220-017-2976-9},
  url = {https://doi.org/10.1007/s00220-017-2976-9}
}

@article{selisko2024dynamical,
      title={Dynamical Mean Field Theory for Real Materials on a Quantum Computer}, 
      author={Johannes Selisko and Maximilian Amsler and Christopher Wever and Yukio Kawashima and Georgy Samsonidze and Rukhsan Ul Haq and Francesco Tacchino and Ivano Tavernelli and Thomas Eckl},
      year={2024},
      eprint={2404.09527},
      archivePrefix={arXiv},
      primaryClass={cond-mat.str-el},
      journal={arXiv preprint},
      url={https://arxiv.org/abs/2404.09527}
}

@article{kemper2024denoising,
  title = {Denoising and Extension of Response Functions in the Time Domain},
  author = {Kemper, Alexander F. and Yang, Chao and Gull, Emanuel},
  journal = {Phys. Rev. Lett.},
  volume = {132},
  issue = {16},
  pages = {160403},
  numpages = {7},
  year = {2024},
  month = {Apr},
  publisher = {American Physical Society},
  doi = {10.1103/PhysRevLett.132.160403},
  url = {https://link.aps.org/doi/10.1103/PhysRevLett.132.160403}
}

@article{ezzell2023dynamical,
  title = {Dynamical decoupling for superconducting qubits: A performance survey},
  author = {Ezzell, Nic and Pokharel, Bibek and Tewala, Lina and Quiroz, Gregory and Lidar, Daniel A.},
  journal = {Phys. Rev. Appl.},
  volume = {20},
  issue = {6},
  pages = {064027},
  numpages = {42},
  year = {2023},
  month = {Dec},
  publisher = {American Physical Society},
  doi = {10.1103/PhysRevApplied.20.064027},
  url = {https://link.aps.org/doi/10.1103/PhysRevApplied.20.064027}
}

@article{hashim2021randomized,
  title = {Randomized Compiling for Scalable Quantum Computing on a Noisy Superconducting Quantum Processor},
  author = {Hashim, Akel and Naik, Ravi K. and Morvan, Alexis and Ville, Jean-Loup and Mitchell, Bradley and Kreikebaum, John Mark and Davis, Marc and Smith, Ethan and Iancu, Costin and O'Brien, Kevin P. and Hincks, Ian and Wallman, Joel J. and Emerson, Joseph and Siddiqi, Irfan},
  journal = {Phys. Rev. X},
  volume = {11},
  issue = {4},
  pages = {041039},
  numpages = {12},
  year = {2021},
  month = {Nov},
  publisher = {American Physical Society},
  doi = {10.1103/PhysRevX.11.041039},
  url = {https://link.aps.org/doi/10.1103/PhysRevX.11.041039}
}

@article{viola1999dynamical,
  title = {Dynamical Decoupling of Open Quantum Systems},
  author = {Viola, Lorenza and Knill, Emanuel and Lloyd, Seth},
  journal = {Phys. Rev. Lett.},
  volume = {82},
  issue = {12},
  pages = {2417--2421},
  numpages = {0},
  year = {1999},
  month = {Mar},
  publisher = {American Physical Society},
  doi = {10.1103/PhysRevLett.82.2417},
  url = {https://doi.org/10.1103/PhysRevLett.82.2417}
}

@misc{qiskit2023research,
    title = {Qiskit Research (v0.0.2)},
    author = {The Qiskit Research developers and contributors},
    year = {2023},
    publisher = {Zenodo},
    doi = {https://doi.org/10.5281/zenodo.7776174},
    URL = {https://github.com/qiskit-community/qiskit-research?tab=readme-ov-file}
}

@article{wallman2016noise,
  title = {Noise tailoring for scalable quantum computation via randomized compiling},
  author = {Wallman, Joel J. and Emerson, Joseph},
  journal = {Phys. Rev. A},
  volume = {94},
  issue = {5},
  pages = {052325},
  numpages = {9},
  year = {2016},
  month = {Nov},
  publisher = {American Physical Society},
  doi = {10.1103/PhysRevA.94.052325},
  url = {https://link.aps.org/doi/10.1103/PhysRevA.94.052325}
}

@article{mukherjee2023ComparativeStudyAdaptive,
  title = {Comparative Study of Adaptive Variational Quantum Eigensolvers for Multi-Orbital Impurity Models},
  author = {Mukherjee, Anirban and Berthusen, Noah F. and Getelina, Jo{\~a}o C. and Orth, Peter P. and Yao, Yong-Xin},
  year = {2023},
  month = jan,
  journal = {Communications Physics},
  volume = {6},
  number = {1},
  pages = {4},
  issn = {2399-3650},
  doi = {10.1038/s42005-022-01089-6},
}

@article{Georges1996DMFT,
  title = {Dynamical mean-field theory of strongly correlated fermion systems and the limit of infinite dimensions},
  author = {Georges, Antoine and Kotliar, Gabriel and Krauth, Werner and Rozenberg, Marcelo J.},
  journal = {Rev. Mod. Phys.},
  volume = {68},
  issue = {1},
  pages = {13--125},
  numpages = {0},
  year = {1996},
  month = {Jan},
  publisher = {American Physical Society},
  doi = {10.1103/RevModPhys.68.13},
  url = {https://link.aps.org/doi/10.1103/RevModPhys.68.13}
}

@article{tikhonov1943stability,
  title={On the stability of inverse problems},
  author={Tikhonov, Andrey Nikolayevich},
  journal={Proceedings of the USSR Academy of Sciences},
  volume={39},
  pages={195--198},
  year={1943}
}

@article{gull2011continuoustimemontecarlo,
  title = {Continuous-time Monte Carlo methods for quantum impurity models},
  author = {Gull, Emanuel and Millis, Andrew J. and Lichtenstein, Alexander I. and Rubtsov, Alexey N. and Troyer, Matthias and Werner, Philipp},
  journal = {Rev. Mod. Phys.},
  volume = {83},
  issue = {2},
  pages = {349--404},
  numpages = {0},
  year = {2011},
  month = {May},
  publisher = {American Physical Society},
  doi = {10.1103/RevModPhys.83.349},
  url = {https://link.aps.org/doi/10.1103/RevModPhys.83.349}
}

@article{mckay2023benchmarking,
  title={Benchmarking quantum processor performance at scale},
  author={McKay, David C and Hincks, Ian and Pritchett, Emily J and Carroll, Malcolm and Govia, Luke CG and Merkel, Seth T},
  journal={arXiv preprint},
  year={2023},
  url={https://doi.org/10.48550/arXiv.2311.05933}
}

@misc{ibmquantumresources,
    note={Quantum processing units, IBM Quantum Platform, https://quantum.ibm.com/services/resources }
}

@article{kokcu2024classification,
  title={Classification of dynamical Lie algebras generated by spin interactions on undirected graphs},
  author={K{\"o}kc{\"u}, Efekan and Wiersema, Roeland and Kemper, Alexander F and Bakalov, Bojko N},
  journal={arXiv preprint},
  year={2024},
  url={https://doi.org/10.48550/arXiv.2409.19797}
}

@article{kokcu2023algebraic,
  title={Algebraic Compression of Free Fermionic Quantum Circuits: Particle Creation, Arbitrary Lattices and Controlled Evolution},
  author={K{\"o}kc{\"u}, Efekan and Camps, Daan and Oftelie, Lindsay Bassman and de Jong, Wibe A and Van Beeumen, Roel and Kemper, AF},
  journal={arXiv preprint},
  year={2025},
  url={https://doi.org/10.48550/arXiv.2303.09538}

}

@article{herbst2022surrogate,
  title = {Surrogate models for quantum spin systems based on reduced-order modeling},
  author = {Herbst, Michael F. and Stamm, Benjamin and Wessel, Stefan and Rizzi, Matteo},
  journal = {Phys. Rev. E},
  volume = {105},
  issue = {4},
  pages = {045303},
  numpages = {13},
  year = {2022},
  month = {Apr},
  publisher = {American Physical Society},
  doi = {10.1103/PhysRevE.105.045303},
  url = {https://link.aps.org/doi/10.1103/PhysRevE.105.045303}
}

@article{besserve2022unraveling,
  title = {Unraveling correlated material properties with noisy quantum computers: Natural orbitalized variational quantum eigensolving of extended impurity models within a slave-boson approach},
  author = {Besserve, P. and Ayral, T.},
  journal = {Phys. Rev. B},
  volume = {105},
  issue = {11},
  pages = {115108},
  numpages = {12},
  year = {2022},
  month = {Mar},
  publisher = {American Physical Society},
  doi = {10.1103/PhysRevB.105.115108},
  url = {https://link.aps.org/doi/10.1103/PhysRevB.105.115108}
}

@article{Nie2024self-consistent,
  title = {Self-Consistent Determination of Single-Impurity Anderson Model Using Hybrid Quantum-Classical Approach on a Spin Quantum Simulator},
  author = {Nie, Xinfang and Zhu, Xuanran and Fan, Yu-ang and Long, Xinyue and Liu, Hongfeng and Huang, Keyi and Xi, Cheng and Che, Liangyu and Zheng, Yuxuan and Feng, Yufang and Yang, Xiaodong and Lu, Dawei},
  journal = {Phys. Rev. Lett.},
  volume = {133},
  issue = {14},
  pages = {140602},
  numpages = {7},
  year = {2024},
  month = {Oct},
  publisher = {American Physical Society},
  doi = {10.1103/PhysRevLett.133.140602},
  url = {https://link.aps.org/doi/10.1103/PhysRevLett.133.140602}
}

@article{capone2007solving,
  title={Solving the dynamical mean-field theory at very low temperatures using the Lanczos exact diagonalization},
  author={Capone, Massimo and de’Medici, Luca and Georges, Antoine},
  journal={Physical Review B—Condensed Matter and Materials Physics},
  volume={76},
  number={24},
  pages={245116},
  year={2007},
  publisher={APS},
  url={https://journals.aps.org/prb/abstract/10.1103/PhysRevB.76.245116}
}

@article{caffarel1994exact,
  title={Exact diagonalization approach to correlated fermions in infinite dimensions: Mott transition and superconductivity},
  author={Caffarel, Michel and Krauth, Werner},
  journal={Physical review letters},
  volume={72},
  number={10},
  pages={1545},
  year={1994},
  publisher={APS}
}

@article{liebsch2011temperature,
  title={Temperature and bath size in exact diagonalization dynamical mean field theory},
  author={Liebsch, Ansgar and Ishida, Hiroshi},
  journal={Journal of Physics: Condensed Matter},
  volume={24},
  number={5},
  pages={053201},
  year={2011},
  publisher={IOP Publishing}
}

@article{schiro2019quantumimpurity,
    author = {Schiro, Marco and Scarlatella, Orazio},
    title = {Quantum impurity models coupled to Markovian and non-Markovian baths},
    journal = {The Journal of Chemical Physics},
    volume = {151},
    number = {4},
    pages = {044102},
    year = {2019},
    month = {07},
    issn = {0021-9606},
    doi = {10.1063/1.5100157},
    url = {https://doi.org/10.1063/1.5100157},
    eprint = {https://pubs.aip.org/aip/jcp/article-pdf/doi/10.1063/1.5100157/15562486/044102\_1\_online.pdf},
}

@article{Scarlatella2021DMFT,
   title={Dynamical Mean-Field Theory for Markovian Open Quantum Many-Body Systems},
   volume={11},
   ISSN={2160-3308},
   url={http://dx.doi.org/10.1103/PhysRevX.11.031018},
   DOI={10.1103/physrevx.11.031018},
   number={3},
   journal={Phys. Rev. X},
   publisher={American Physical Society (APS)},
   author={Scarlatella, Orazio and Clerk, Aashish A. and Fazio, Rosario and Schiró, Marco},
   year={2021},
   month=jul}

\clearpage
\onecolumngrid
\appendix

\renewcommand\thefigure{S\arabic{figure}}  
\renewcommand\thetable{S\arabic{table}}  
\setcounter{figure}{0}

\section{Extended discussion of DMFT}\label{asec: DMFT details}
Using DMFT, the correlated behavior of the lattice model that describes a real material is related to an impurity model. In our study, we performed DMFT for the half-filled Hubbard model in the Bethe lattice (see~\cref{subsec: SGS DMFT}) using an approximate impurity ground state represented by a sum of Gaussian states. The Hamiltonian that describes the $N$-orbital Hubbard model is given by~\cref{aeq: Hubbard Hamiltonian}, where $h$ is the hopping strength, $\mu$ is the chemical potential, and $U$ is the Coulomb interaction strength.

\begin{align}\label{aeq: Hubbard Hamiltonian}
    \widehat{\mbf{H}}_{\text{Hubbard}} &= \widehat{\mbf{H}}_0 + \widehat{\mbf{H}}_{\text{int}} \\
    \widehat{\mbf{H}}_0 &= -\sum^{N}_{\langle i,j\rangle=1,\sigma} h(\hat{c}^\dag_{i\sigma}\hat{c}^\phant_{j\sigma}+\text{h.c.}) \nonumber -\sum^{N}_{i=1,\sigma}\mu \hat{c}^\dag_{i\sigma}\hat{c}^\phant_{i\sigma} \\
    \widehat{\mbf{H}}_{\text{int}} &= \sum^{N}_{i}  U \hat{c}^\dag_{i\uparrow}\hat{c}^\phant_{i\uparrow}\hat{c}^\dag_{i\downarrow}\hat{c}^\phant_{i\downarrow}
\end{align}

\noindent
Here, the brackets in $\langle i,j\rangle$ indicate that the hopping terms exist only between nearest neighbors. A notable distinction between the Hubbard model and the impurity model lies in the $\widehat{\mbf{H}}_\text{int}$ term, which sums over all orbitals instead of being restricted to the impurity orbitals. While self-consistent embeddings exist for impurity clusters (e.g. the so-called cellular DMFT) and our methods do allow for implementation with more than one impurity orbital, we limit the scope of this work to the single impurity Anderson model. Consequently, all further equations presented will consider $N_I=1$. 

A key quantity required to enforce DMFT self-consistency is the impurity GF, $G^R_\text{imp}$ given by~\cref{aeq: impurity GF} in its (a) time dependent, (b) real, and (c) Matsubara (imaginary) frequency-dependent forms ($\omega_n = \frac{(2n + 1)\pi}{\beta}$ for fermions and $\beta$ is inverse temperature).

\begin{subequations}\label{aeq: impurity GF}
\begin{align}
    G^R_\text{imp}(t) &= -i\langle \Psi^{(0)}|\{\hat{d}^\phant_\sigma(t),\hat{d}^\dag_\sigma\}|\Psi^{(0)}\rangle \label{aeq: time impurity GF}\\
    G^R_\text{imp}(\omega) &= \langle\Psi^{(0)}|\hat{d}^\phant_\sigma [(\omega+E^{(0)}+i\eta)\mbf{I}-\widehat{\mbf{H}}_\text{imp}]^{-1}\hat{d}^\dag_\sigma|\Psi^{(0)}\rangle + \langle\Psi^{(0)}|\hat{d}^\dag_\sigma [(\omega-E^{(0)}+i\eta)\mbf{I}+\widehat{\mbf{H}}_\text{imp}]^{-1}\hat{d}^\phant_\sigma|\Psi^{(0)}\rangle \label{aeq: freq imp GF} \\
    \mcG_\text{imp}(i\omega_n) &= \int_{-\infty}^\infty \frac{d\omega}{\pi} \frac{-\text{Im}[G^R_\text{imp}(\omega)]}{i\omega_n-\omega} \label{aeq: Matsubara imp GF transfo}
\end{align}
\end{subequations}

\noindent
In~\cref{aeq: freq imp GF}, a small, positive constant $\eta$ is added to ensure convergence of the Fourier transform from the time dependent form in \eqref{aeq: time impurity GF}.

Each of these forms is equivalent in terms of their dynamical information. However, in practice, the Matsubara GF tends to be smooth, while the time dependent and frequency-dependent impurity GFs exhibit long-lived oscillations and sharp peaks, respectively. To ensure stable convergence to the self-consistent bath parameters during the numerical fitting procedures used in DMFT, it is most effective to work with the impurity GF on the Matsubara frequency axis. Although the forms mentioned refer to operators that exist within an exponential Hilbert space,~\cref{aeq: Matsubara imp GF transfo} can also be expressed in a manner that separates the noninteracting component of the impurity GF --- which includes the impurity on-site energy ($\nu_{11}=\epsilon_i-\mu$, with $\mu$ the chemical potential) and impurity-bath hybridization ($\Delta(i\omega_n)$) --- and the local electron interactions of the impurity orbital, called the self energy ($\Sigma_\text{imp}(i\omega_n)$).

\begin{align}\label{aeq: Matsubara impurity GF}
    \mcG_\text{imp}(i\omega_n) &= [i\omega_n - \epsilon_i + \mu - \Delta(i\omega_n) -\Sigma_\text{imp} (i\omega_n)]^{-1}
\end{align}

The hybridization with the bath has an exact analytical form of

\begin{align}\label{aeq: hybridization}
    \Delta(i\omega_n)=\sum_{b=1}^{N_B} \frac{|V_b|^2}{i\omega_n-\epsilon_b}
\end{align}

\noindent
where $V_b, \epsilon_b$ are the bath hoppings and on-site energies. The self-energy of the impurity model comes from Dyson's equation

\begin{align}\label{aeq: dyson's eqn}
    \mbf{G} &= \mbf{G}^0 + \mbf{G}^0 \mbf{\Sigma} \mbf{G}
\end{align}

\noindent
where $\mbf{G}^0$ is the non-interacting GF (which has an exact analytical form) and $\mbf{\Sigma}$ is the self-energy. Here, the quantities in~\cref{aeq: dyson's eqn} are $2N\times 2N$ matrices, with each of the elements signifying all of the system's single-particle GFs with different spin and orbital creation/annihilation operators [$\hat{d}^{(\dag)}_\sigma,\hat{c}^{(\dag)}_{1\sigma},...,\hat{c}^{(\dag)}_{N_B\sigma}$].

As discussed in the main text, the impurity model is connected to the lattice model through its self-energy. In the case of the lattice model with infinite coordination, the lattice self-energy ($\Sigma_\text{latt}$) for each interacting orbital is independent of crystal momentum $\mbf{k}$. This can be understood as each individual orbital being influenced by many-body interactions from the bulk in a uniform manner, regardless of its actual position within the lattice, due to the infinite coordination.

\begin{align}\label{aeq: local self energy}
    \Sigma_{\text{latt}}(i\omega_n,\mbf{k})\rightarrow \Sigma_{\text{latt,loc}}(i\omega_n)
\end{align}

\noindent
Using the self-consistent condition

\begin{align}\label{aeq: self consistent}
    \Sigma_\text{imp} &\approx \Sigma_{\text{latt,loc}}
\end{align}

\noindent
the approximation of the lattice GF at the current iteration of DMFT can be evaluated with~\cref{aeq: lattice GF} using the assumption made with \eqref{aeq: self consistent}. 

\begin{align}\label{aeq: lattice GF}
    G_\text{latt,loc}(i\omega_n) &\approx \int_{-\infty}^{\infty} dx \frac{\rho(x)}{i\omega_n-x+\mu-\Sigma_\text{imp}(i\omega_n)}    \\
    \rho(x) &= \frac{\sqrt{4h^2-x^2}}{2\pi h^2}
\end{align}

\noindent
where $\rho(x)$ is the density of states for a Bethe lattice and $h$ is the hopping strength of the Hubbard Hamiltonian. Returning to~\cref{aeq: Matsubara impurity GF}, we can obtain new bath parameters for the impurity model by minimizing

\begin{align}\label{aeq: hybridization fitting}
    \mathcal{L}(V_b,\epsilon_b)=\left|\Delta(i\omega_n) - \left[\frac{1}{G_\text{latt,loc}(i\omega_n)} - i\omega_n + \epsilon_i - \mu + \Sigma_\text{imp}(i\omega_n)\right]\right|^2.
\end{align}

\noindent
It is this fitting procedure that necessitates the dynamical quantities ($\mcG_\text{imp}, \Sigma_\text{imp}, G_\text{latt,loc}, \Delta$) to vary smoothly with respect to their dependent variable, motivating the choice to perform DMFT using Matsubara frequencies. After some iterations of DMFT, if the change in $V_b,\epsilon_b$ is below some threshold, the impurity model's bath can be considered self-consistent in terms of describing the many-body effects on the atomic orbitals of the lattice model, thus concluding the DMFT loop.

\section{Gaussian states}\label{asec: Gaussian states}

A fermionic Gaussian Hamiltonian (FGH) for an $N$-orbital system has the form of~\cref{aeq: Gaussian Hamiltonian}, where $\mbf{H}^{\sigma\sigma'}_{ij}$ will be referred to as elements of the \textit{hopping matrix}.

\begin{align}\label{aeq: Gaussian Hamiltonian}
    \widehat{\mbf{H}}_0 &= \sum^{N}_{ij,\sigma\sigma'} \mbf{H}^{\sigma\sigma'}_{ij} \hat{c}^\dagger_{i\sigma}\hat{c}_{j\sigma'}
\end{align}

\noindent
with $\sigma,\sigma'\in\{\uparrow,\downarrow\}$. While $\widehat{\mbf{H}}_0$ is a linear operator that exists in the full Hilbert space ($2^{2N}\times 2^{2N}$), the hopping matrix $\mbf{H}$ of size $2N\times 2N$ contains all of the information needed to characterize eigenstates of the FGH. 

All fermionic Gaussian states (FGS) $\ket{\phi}$ can be described by their set of occupations within a skew-symmetric covariance matrix $\mbf{M}$ that takes the form

\begin{align}\label{aeq: covariance matrix}
[\mbf{M}]_{pq} &= -\frac{i}{2}\bra{\phi}[\hat{\gamma}_p,\hat{\gamma}_q]\ket{\phi}
\end{align}

\noindent
where $[\hat{\gamma}_p,\hat{\gamma}_q]$ is the commutator of Majorana operators.  

The magnitude of the overlap between two FGS can be computed with 

\begin{align}\label{aeq: overlap}
    \langle \phi_i | \phi_j \rangle = \sqrt{\text{Pf}\left(\frac{\mbf{M}_i + \mbf{M}_j}{2}\right)}
\end{align}

\noindent
where $\text{Pf}(.)$ is the Pfaffian. We assume $\langle \phi_i | \phi_j \rangle$ to always be positive, which can be guaranteed by gauge-fixing our states $\ket{\phi_k}$ on the level of a Fock basis element~\cite{boutin2021quantum}.

The matrix elements of any monomial of Majorana operators can be computed with~\cref{aeq: matrix element}, where $\bm{\hat{\gamma}}[\mbf{x}]=\hat{\gamma}_1^{x_1}\hat{\gamma}_2^{x_2}...\hat{\gamma}_{2N}^{x_{2N}}$ and $x_k\in\{0,1\}$.

\begin{align}\label{aeq: matrix element}
    \langle \phi_i | \bm{\hat{\gamma}}[\mbf{x}] | \phi_j \rangle = \langle \phi_i | \phi_j \rangle \text{Pf}(i\mbf{\Delta}[\mbf{x}]^*)
\end{align}

\noindent
Here, $\mbf{\Delta}$ is a skew-symmetric matrix described by

\begin{align}\label{aeq: delta}
    \mbf{\Delta} = (-2\mbf{I}+i(\mbf{M}_i-\mbf{M}_j))(\mbf{M}_i+\mbf{M}_j)^{-1}
\end{align}

\noindent
and $\mbf{\Delta}[\mbf{x}]^*$ represents the submatrix with rows and columns supported by $\mbf{x}$. $\mbf{\Delta}^*$ denotes the complex conjugate. Eqs. \eqref{aeq: overlap} and \eqref{aeq: matrix element} are readily evaluated with all the information obtained from the hopping matrix, which scales linearly with the system size, allowing for efficient calculations using a basis of non-orthogonal FGS.

\section{Correlation Function and Corresponding Circuits}\label{app: partial compression details}

The correlation function we calculate (for a single impurity) is given as follows:
\begin{align}\label{aeq:GF}
    G_\text{imp}^R(t) = -i\bra{\Psi^{(0)}} \{ \hat{d}(t), \hat{d}^\dagger \} \ket{\Psi^{(0)}}.
\end{align}
Here $\hat{d}(t)$ represents the time evolved $\hat{d}$ operator in the Heisenberg picture, the state $\ket{\Psi^{(0)}}$ is the ground state of the impurity model, and $\hat{d}$ and $\hat{d}^\dagger$ are fermion annihilation and creation operators (for now, we neglect the spin indices $\sigma$), which are not unitary and cannot be directly implemented in a quantum circuit. However, the linear combinations $\hat{\gamma}_+ = \hat{d} +\hat{d}^\dagger$ and $\hat{\gamma}_- = i(\hat{d} -\hat{d}^\dagger)$  are unitary operators, and therefore can be applied on a quantum circuit. The Green's function in \eqref{aeq:GF} can then be written in terms of $\hat{\gamma}_\pm$ as follows
\begin{align}
\begin{split}
    G^R_\text{imp}(t) &= \frac{1}{4}\bra{\Psi^{(0)}} \{ (\hat{\gamma}_+ + i \hat{\gamma}_-)(t),(\hat{\gamma}_+ - i\hat{\gamma}_-) \}  \ket{\Psi^{(0)}} \\
    &= \frac{1}{4}\bra{\Psi^{(0)}} \{ \hat{\gamma}_+(t), \hat{\gamma}_+ \} \ket{\Psi^{(0)}} - \frac{i}{4}\bra{\Psi^{(0)}} \{\hat{\gamma}_+(t), \hat{\gamma}_-\}  \ket{\Psi^{(0)}} \\
    & \hspace{4mm} + \frac{i}{4}\bra{\Psi^{(0)}}  \{\hat{\gamma}_-(t), \hat{\gamma}_+\}  \ket{\Psi^{(0)}} + \frac{1}{4}\bra{\Psi^{(0)}} \{ \hat{\gamma}_-(t), \hat{\gamma}_-  \} \ket{\Psi^{(0)}} \\
\end{split}
\end{align}
In addition to their unitarity, $\gamma_{\pm}$ are also Hermitian operators, which yields
\begin{align}
\begin{split}
    G^R_\text{imp}(t)
    &= \frac{1}{4}\mathrm{Re} \bra{\Psi^{(0)}} \hat{\gamma}_+(t) \hat{\gamma}_+ \ket{\Psi^{(0)}} - \frac{i}{4}\mathrm{Re}\bra{\Psi^{(0)}} \hat{\gamma}_+(t) \hat{\gamma}_- \ket{\Psi^{(0)}} \\
    & \hspace{4mm} + \frac{i}{4}\mathrm{Re}\bra{\Psi^{(0)}}  \hat{\gamma}_-(t) \hat{\gamma}_+ \ket{\Psi^{(0)}} + \frac{1}{4}\mathrm{Re}\bra{\Psi^{(0)}}\hat{\gamma}_-(t) \hat{\gamma}_-  \ket{\Psi^{(0)}} \\
    &= \frac{1}{4}\sum_{\substack{a,b \in \{-,+\}}} s_{a,b} \: \mathrm{Re} \bra{\Psi^{(0)}}  \hat{\gamma}_a(t) \hat{\gamma}_b  \ket{\Psi^{(0)}},
\end{split}
\end{align}
where we defined $s_{++} = 1$, $s_{+-} = -i$, $s_{-+} = i$, $s_{--} = 1$.
To calculate each term on a quantum computer, we approximate the ground state as a linear combination of the FGS:
\begin{align}
    \ket{\psi^{(0)}} = \sum_{k=1}^\chi \alpha^{(0)}_k \ket{\phi_k} = \sum_{k=1}^\chi \alpha^{(0)}_k U_k \ket{0}.
\end{align}
Here $\ket{\phi_k}$ are FGS, and therefore $U_k$ can be chosen as free fermion evolution operators~\cite{kokcu2023algebraic}.
With this, the Green's function can be calculated as follows:

\begin{align}
    G^R_\text{imp}(t) = \frac{1}{4}\sum_{\substack{a,b \in \{-,+\}}} s_{ab} \mathrm{Re} \left(\sum_{i,j = 1}^\chi \alpha_i^* \alpha_j \bra{0} U_i^\dagger e^{itH}\hat{\gamma}_a e^{-itH} \hat{\gamma}_b  U_j  \ket{0} \right),
\end{align}
where we applied the Heisenberg operator evolution $\hat{\gamma}_a(t) = e^{itH}\hat{\gamma}_a e^{-itH}$ with $\mbf{\widehat{H}}_\text{imp}$ being simplified to a general Hamiltonian $H$ --- assumed to have a limited number of interaction terms --- for notational convenience.

We use the Hadamard test to calculate each term using a quantum computer. This requires us to use an ancilla qubit in the $\ket{+}$ state, apply the unitary $ U_i^\dagger \hat{\gamma}_a(t) \hat{\gamma}_b  U_j$ in a controlled manner i.e. only for the component that has the ancilla qubit in $\ket{1}$ state, and finally measure the ancilla qubit on $X$ direction for the real part, and $Y$ direction for the imaginary part of the transition amplitude. This corresponds to the following circuit:
\begin{align}\label{eq:bare_circuit}
    \vcenter{\hbox{\includegraphics[height = 0.16\columnwidth]{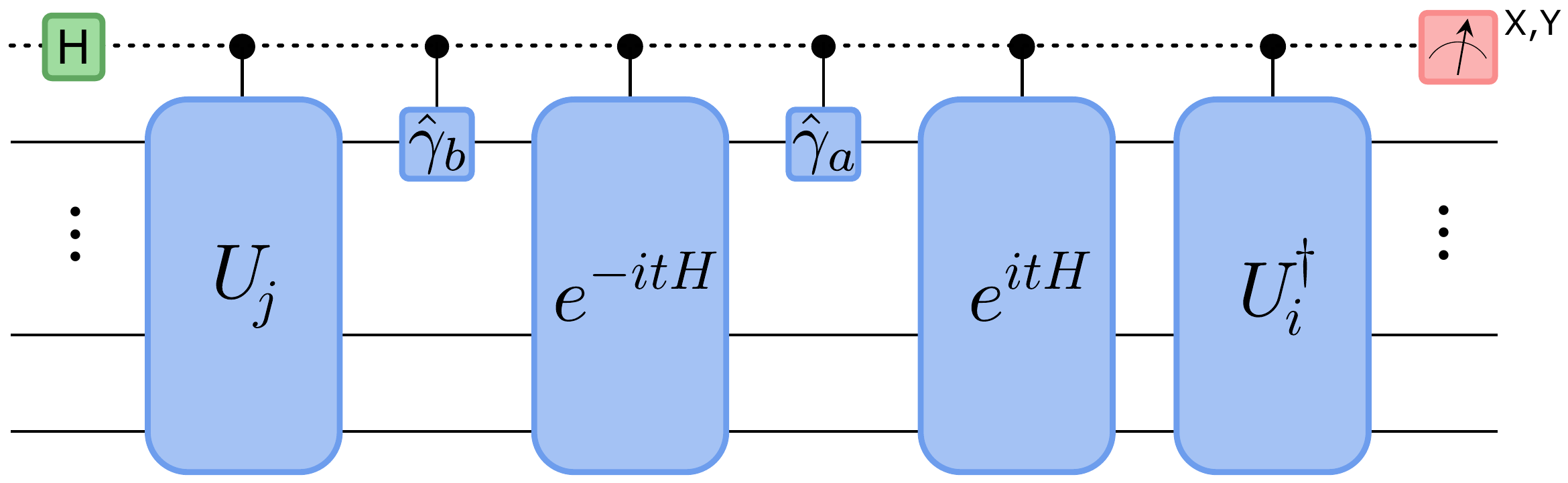}}}
\end{align}
\noindent
We will apply several layers of simplification to this circuit, which will be explained in detail in the following subsections. 

\subsection{Controlled unitary simplification}\label{app:controlled_unitary_simplification}
First, we will show how to simplify the circuit given in \eqref{eq:bare_circuit} by half in depth for \textit{any} Hamiltonian $H$, and \textit{any} unitaries $U_i$, $U_j$, $\hat{\gamma}_a$ and $\hat{\gamma}_b$. We will frequently use the following equality:
\begin{align}\label{eq:lost_control}
    \vcenter{\hbox{\includegraphics[width = 0.65\columnwidth]{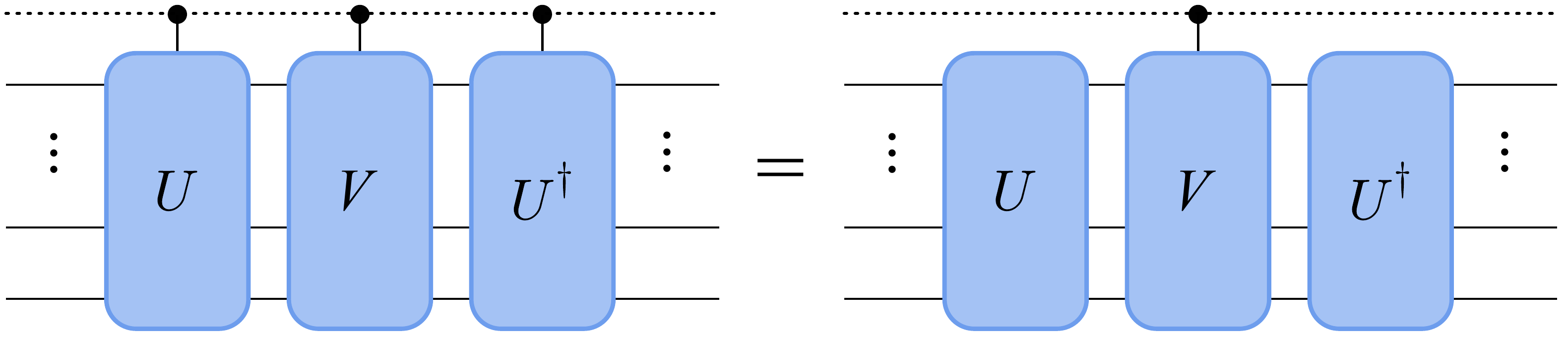}}}
\end{align}
To see that these two circuits are equal, let us follow what they implement for each state of the ancilla qubit. If the ancilla is in the $\ket{0}$ state, none of the unitaries are applied in the circuit on the left since they are controlled, and therefore the operation is the identity matrix $I$. In the circuit on the right, first $U$ is applied. $V$ is not applied since it is controlled, and then $U^\dagger$ is applied, yielding that the circuit is equivalent to $U^\dagger U = I$, which is the same as the circuit on the left-hand side. If the ancilla is in state $\ket{1}$, it can be seen that both circuits will apply $U^\dagger V U$, and therefore they are equal in this case as well.

A direct application of \eqref{eq:lost_control} on the circuit in \eqref{eq:bare_circuit} allows us to remove the controls on the time evolution operators by setting $V \equiv \hat{\gamma}_a$, and $U \equiv e^{-i Ht}$, which yields
\begin{align}\label{eq:circ2}
    \vcenter{\hbox{\includegraphics[height = 0.16\columnwidth]{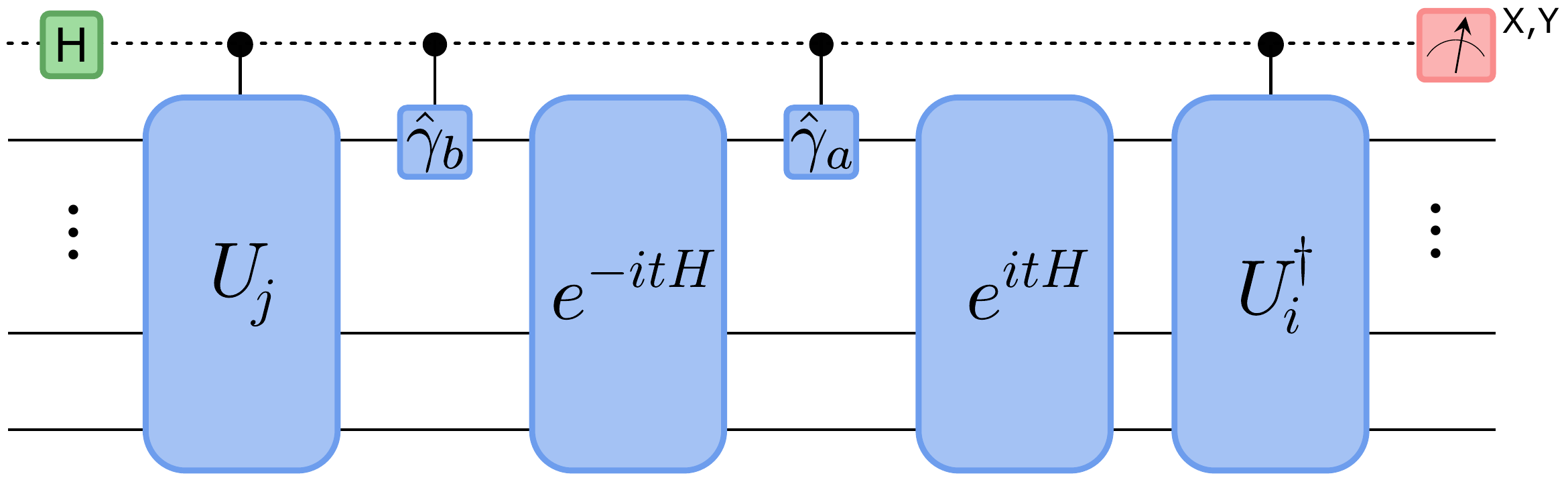}}}
\end{align}
To make use of \eqref{eq:lost_control} further, let us introduce an identity operator by using controlled $U_i$ and controlled $U_i^\dagger$ as follows:
\begin{align}\label{eq:circ3}
    \vcenter{\hbox{\includegraphics[height = 0.2\columnwidth]{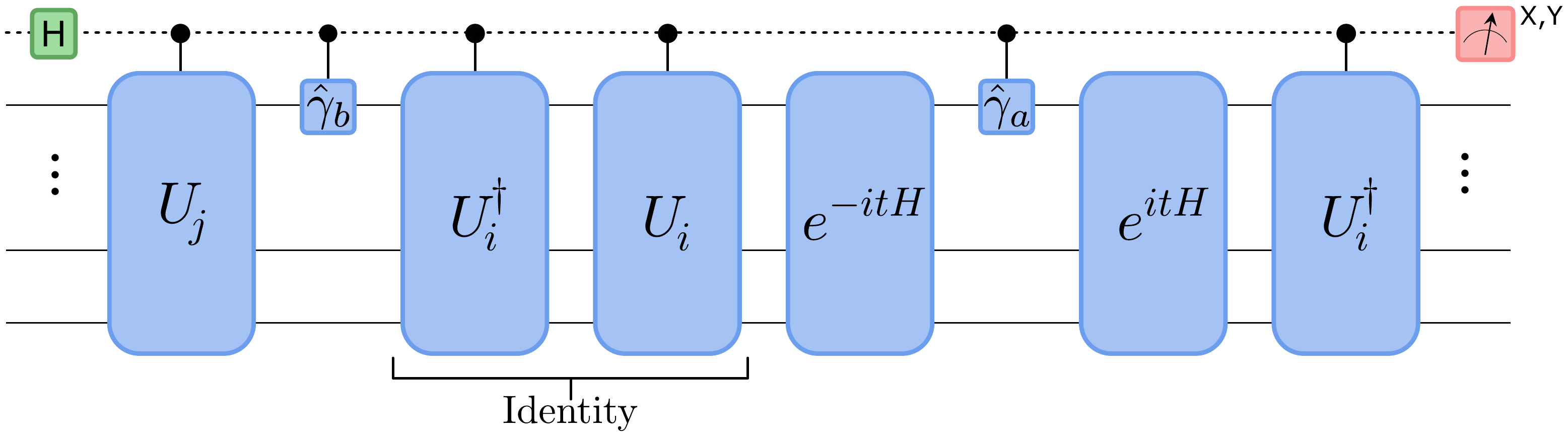}}}
\end{align}
We can lose the controls on the $U_i$ we placed and the $U_i^\dagger$ at the end by using \eqref{eq:lost_control}:
\begin{align}\label{eq:circ4}
    \vcenter{\hbox{\includegraphics[height = 0.2\columnwidth]{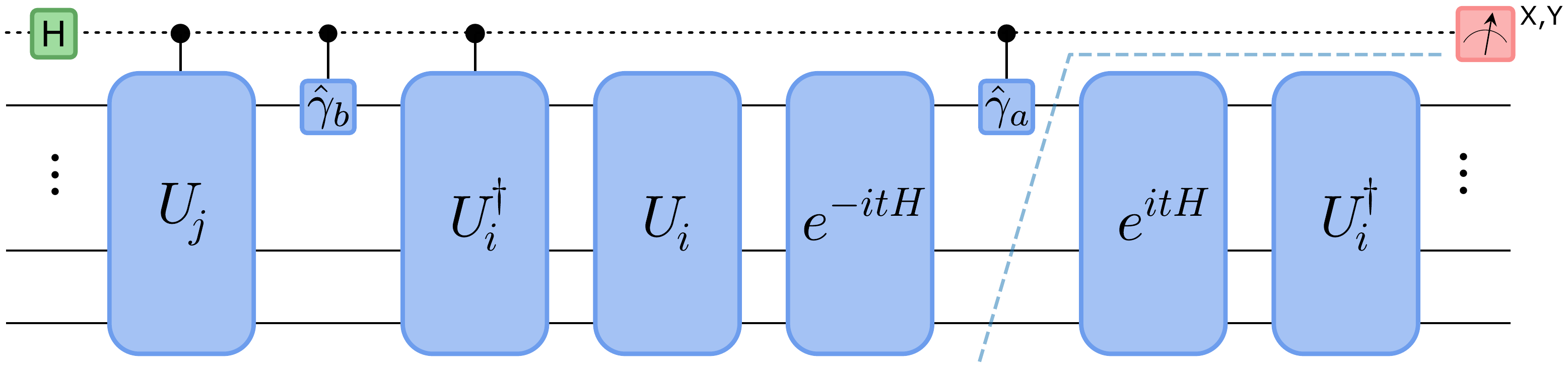}}}
\end{align}
Now it can be seen that $U_i^\dagger$ and $e^{itH}$ at the end are outside the light cone of the ancilla measurement, and therefore can be discarded. In addition, the operators $U_i$ and controlled $U_i^\dagger$ are equivalent to an anti-controlled $U_i$, i.e., it is $U_i$ when the ancilla is in $\ket{0}$, and it is the identity when the ancilla is in $\ket{1}$. Applying these changes, we arrive at the following:
\begin{align}\label{eq:circ5}
    \vcenter{\hbox{\includegraphics[height = 0.16\columnwidth]{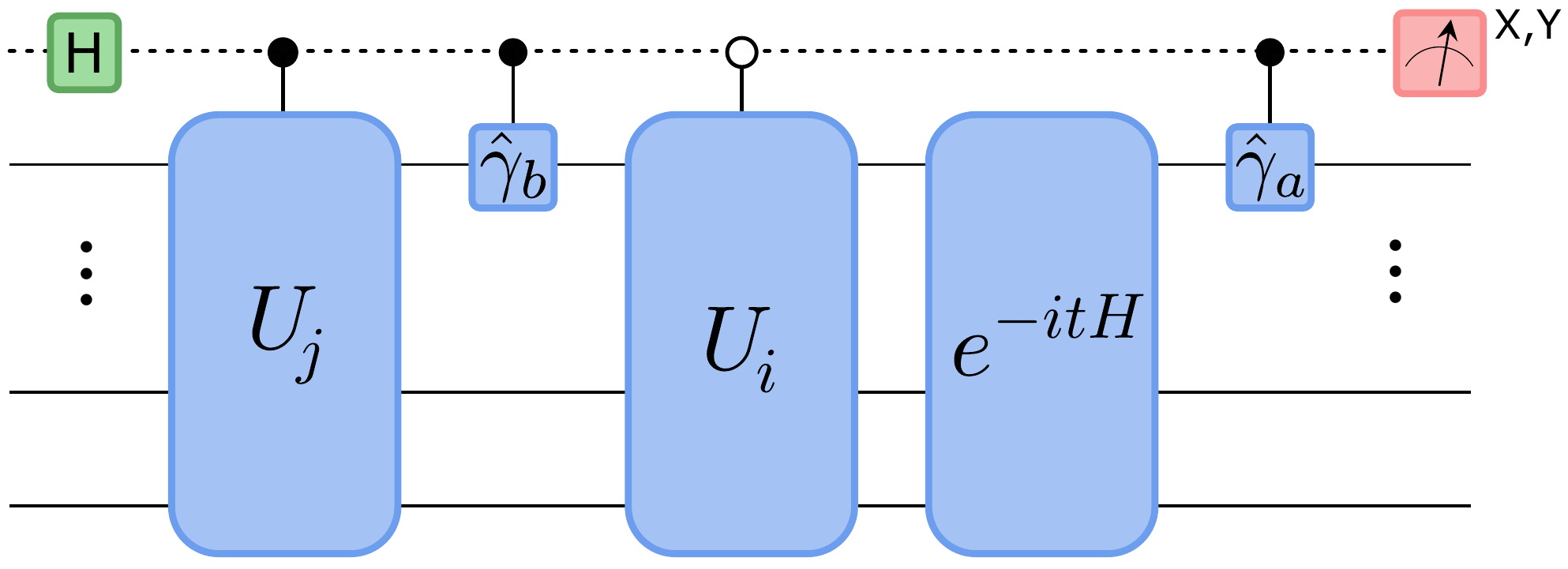}}}
\end{align}
Finally, anti-controlled $U_i$ and controlled $\hat{\gamma}_b$ act on separate parts of the Hilbert space, and therefore commute. We can use this to change their order, to obtain the following simplified circuit for calculating the transition amplitude $\bra{0} U_i^\dagger \hat{\gamma}_a(t) \hat{\gamma}_b  U_j  \ket{0}$:
\begin{align}\label{eq:simple_circ}
    \vcenter{\hbox{\includegraphics[height = 0.2\columnwidth]{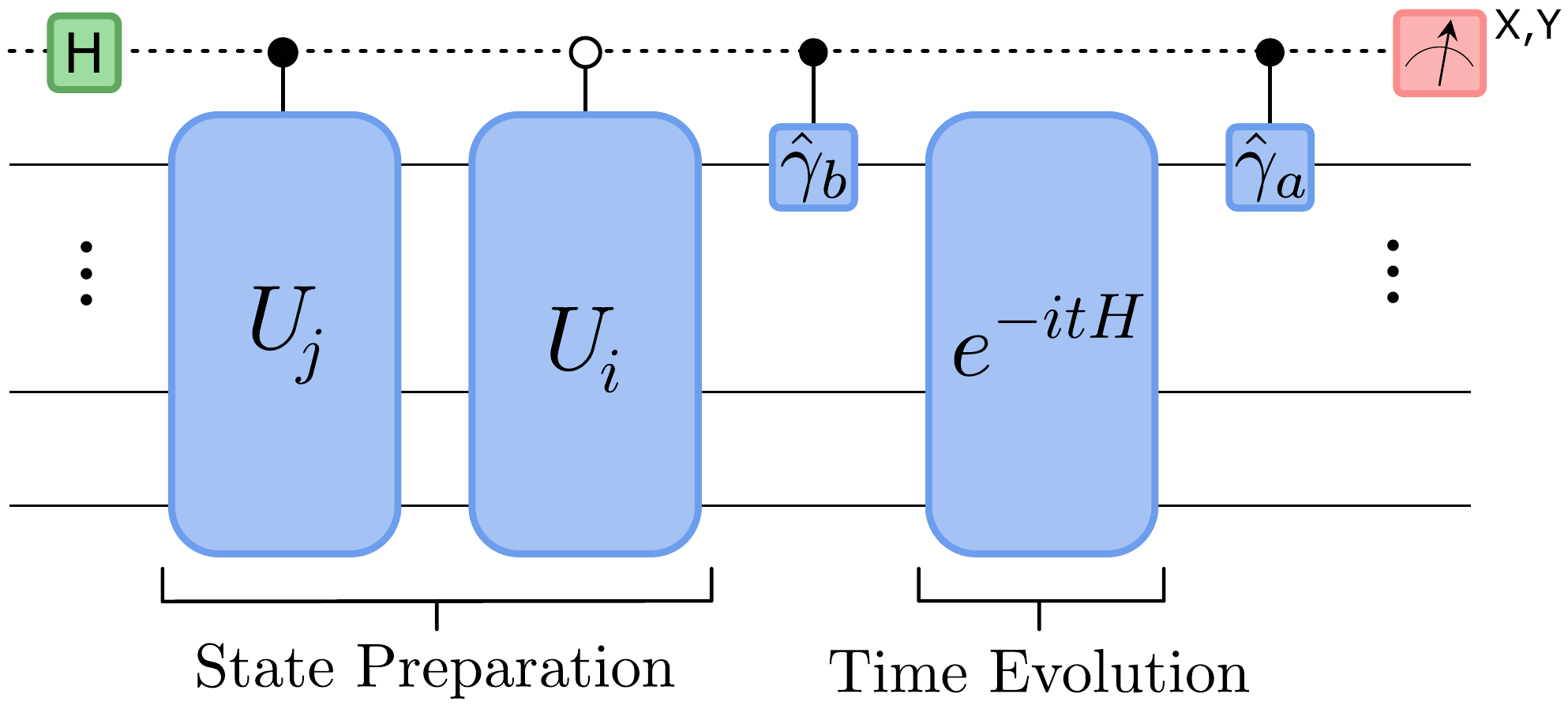}}}.
\end{align}
Grouping the controlled $U_j$ and anti-controlled $U_i$ together allows us to separate the circuit into state preparation and time evolution circuits.

\subsection{Partial compression of the second-order Trotter time evolution}

In this subsection, we will focus on the time evolution operator $e^{-it \mbf{\widehat{H}}_\text{imp}}$ for the impurity Hamiltonian given in Eq. \eqref{eq: impurity hamiltonian}. Let us separate the Hamiltonian into two parts as follows:
\begin{align}
    \widehat{\mbf{H}}_\text{imp} = \widehat{\mbf{H}}_2 + \widehat{\mbf{H}}_4,
\end{align}
where
\begin{align}
    \widehat{\mbf{H}}_2 = \sum_{ij\sigma}^{N_I} \nu_{ij}\hat{d}^\dagger_{i\sigma}\hat{d}^\phant_{j\sigma} + \sum_{i}^{N_I} \sum_{b\sigma}^{N_B} \epsilon_{ib} \hat{c}^\dagger_{ib\sigma}\hat{c}^\phant_{ib\sigma} + \sum_{i}^{N_I} \sum^{N_B}_{b\sigma}V^i_b (\hat{d}^\dagger_{i\sigma}\hat{c}^\phant_{ib\sigma} + \text{h.c.})
\end{align}
which contains all the quadratic terms that are free fermionic, and 
\begin{align}\label{aeq:quartic}
    \widehat{\mbf{H}}_4 = U \sum^{N_I}_{i}
    \hat{n}_{i\uparrow} \hat{n}_{i\downarrow} 
    +U' \sum^{N_I}_{i\neq j}\sum_{\sigma\sigma'}
    \hat{n}_{i\sigma} \hat{n}_{j\sigma'}    ,
\end{align}
with $\hat{n}_{i\sigma}=\hat{d}^\dagger_{i\sigma}\hat{d}^\phant_{i\sigma}$. This contains the quartic terms that correspond to the Coulomb interactions taking place on the impurity orbitals. For clarity, $\widehat{\mbf{H}}_4$ contains only one term for a single impurity model ($\widehat{\mbf{H}}_4(N_I=1)=U\hat{n}_\uparrow\hat{n}_\downarrow$). 

To implement $e^{-it \widehat{\mbf{H}}_\text{imp}}$, we will use the second-order Trotter-Suzuki formula:
\begin{align}\label{eq:trotterization}
    e^{-it \widehat{\mbf{H}}_\text{imp} }= \left( e^{-i\frac{t}{2r} 
    \widehat{\mbf{H}}_2}e^{-i\frac{t}{r}  \widehat{\mbf{H}}_4}e^{-i\frac{t}{2r}  \widehat{\mbf{H}}_2}\right)^{r} + \mathcal{O} \left \{ \left( \Big| \Big| \big[\widehat{\mbf{H}}_2,[\widehat{\mbf{H}}_2,\widehat{\mbf{H}}_4] \big] \Big|\Big| + \Big| \Big| \big[\widehat{\mbf{H}}_4,[\widehat{\mbf{H}}_2,\widehat{\mbf{H}}_4] \big] \Big|\Big| \right) \frac{t^3}{r^2} \right \}  ,
\end{align}
where $||.||$ refers to the spectral norm. We would like to note that in the generic case, this spectral norm would be $\mathcal{O}(N_I^4)$ simply because the number of Pauli strings in $\widehat{\mbf{H}}_4$ after the Jordan-Wigner transformation is proportional to the square of the number of impurity orbitals ($U'$ term in \cref{aeq:quartic} is a sum over impurity pairs). 
This then yields that the Trotter error given in \eqref{eq:trotterization} is $\mathcal{O}(N_I^4 t^3 r^{-2})$, and is independent of $N_B$. For simplicity, we again set $N_I = 1$ and $N_B=3$ for the following illustrations.

Let us now generate a circuit for the trotterized time evolution given in \eqref{eq:trotterization}.
$\widehat{\mbf{H}}_2$ contains on-site terms on the bath orbitals and the impurity orbitals, as well as hoppings between the impurity orbitals and each bath orbital. However, each of these terms is limited within its spin sector: i.e., there is no hopping altering the spin of the electrons. Thus, $\widehat{\mbf{H}}_2$ is actually two free fermionic Hamiltonians that act on effectively different orbitals, which are represented by different sets of qubits. Using this information, and  the algebraic compression of free fermions on a generic lattice given in~\cite{kokcu2023algebraic}, we can then build the following circuit for a single Trotter step ($r=1$) given in \eqref{eq:trotterization}:

\begin{align}\label{eq:Trot_circ}
    \vcenter{\hbox{\includegraphics[height = 0.205\columnwidth]{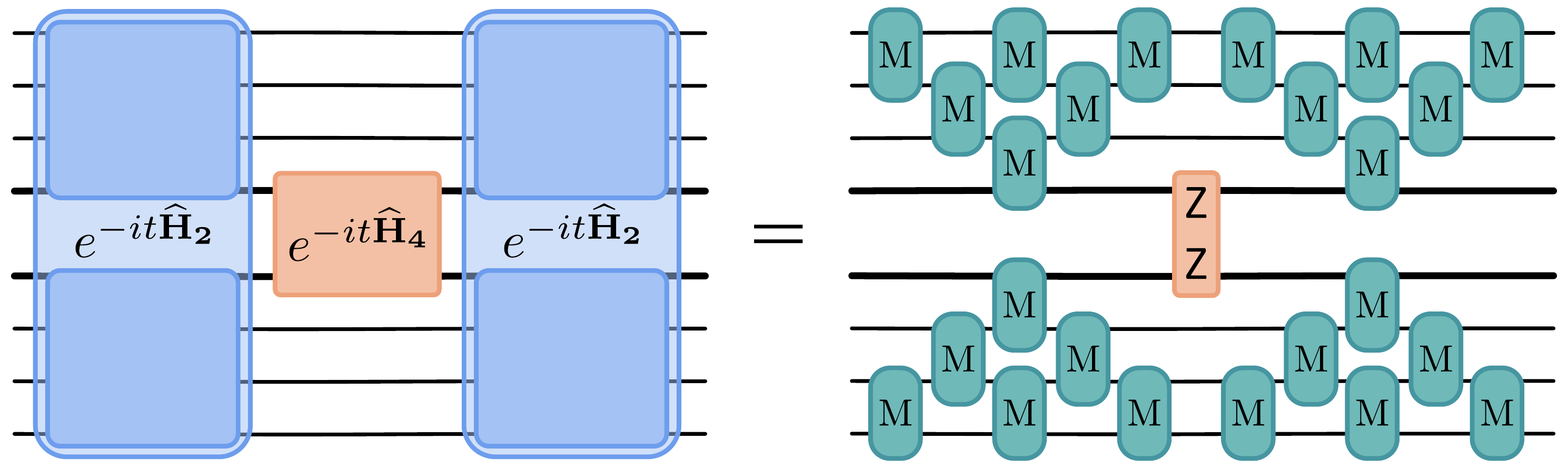}}}
\end{align}
where we used the fact that the interaction term in $\widehat{\mbf{H}}_4$ is a density-density interaction which becomes a $ZZ$ rotation after a Jordan-Wigner transformation. The bold qubit wires correspond to the impurity qubits, whereas the other thinner wires are the bath qubits. Note that we compressed the spin $\downarrow$ hoppings in the lower register into a triangle, and the spin $\uparrow$ hoppings in the upper register into an upside-down triangle. This is possible since the block rules are symmetric under the up-down parity~\cite{kokcu2022algebraic,camps2022algebraic, kokcu2023algebraic}.

Now let us compress one Trotter step partially. For certain free fermion matchgates, or TFXY blocks, we can use the block properties to reduce a Trotter step as follows:
\begin{align}\label{eq:trotter_step_compression}
    \vcenter{\hbox{\includegraphics[height = 0.18\columnwidth]{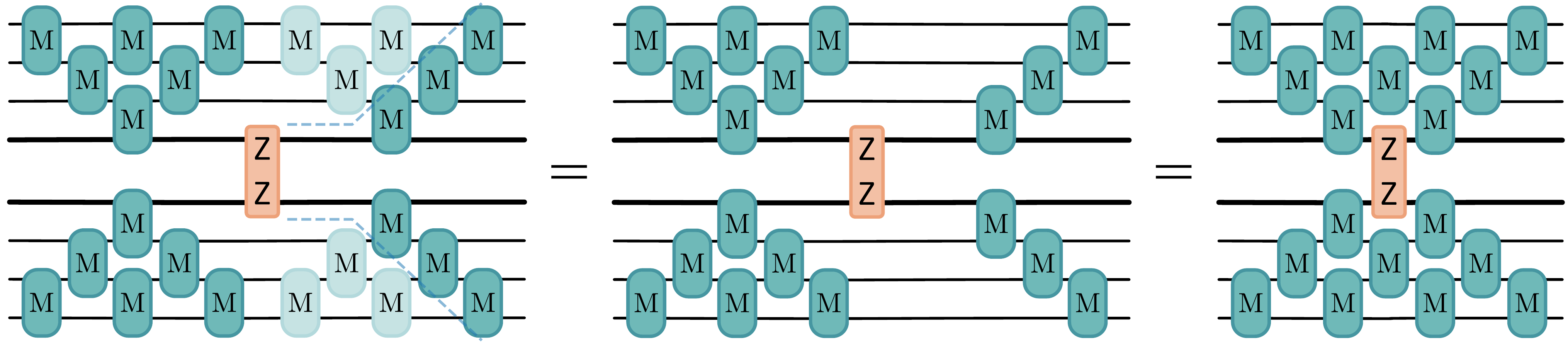}}}
\end{align}
The lighter-colored blocks on the left-most circuit act only on the bath qubits, and therefore can commute with the impurity term, which only acts on the impurity qubits. This allows the triangles on the left side of the $ZZ$ gate to absorb them via the compression given in Thm. 1, Ref.~\cite{kokcu2022algebraic}, which leads to the circuit in the middle. Note that this cuts down the number of matchgates approximately by half for each Trotter step.

A similar partial compression can be applied to the subsequent Trotter steps as well ($r>1$), which leads to an even greater simplification for the time evolution circuit. As an example, let us consider $r=3$:
\begin{align}\label{eq:3trottersteps}
    \vcenter{\hbox{\includegraphics[height = 0.2\columnwidth]{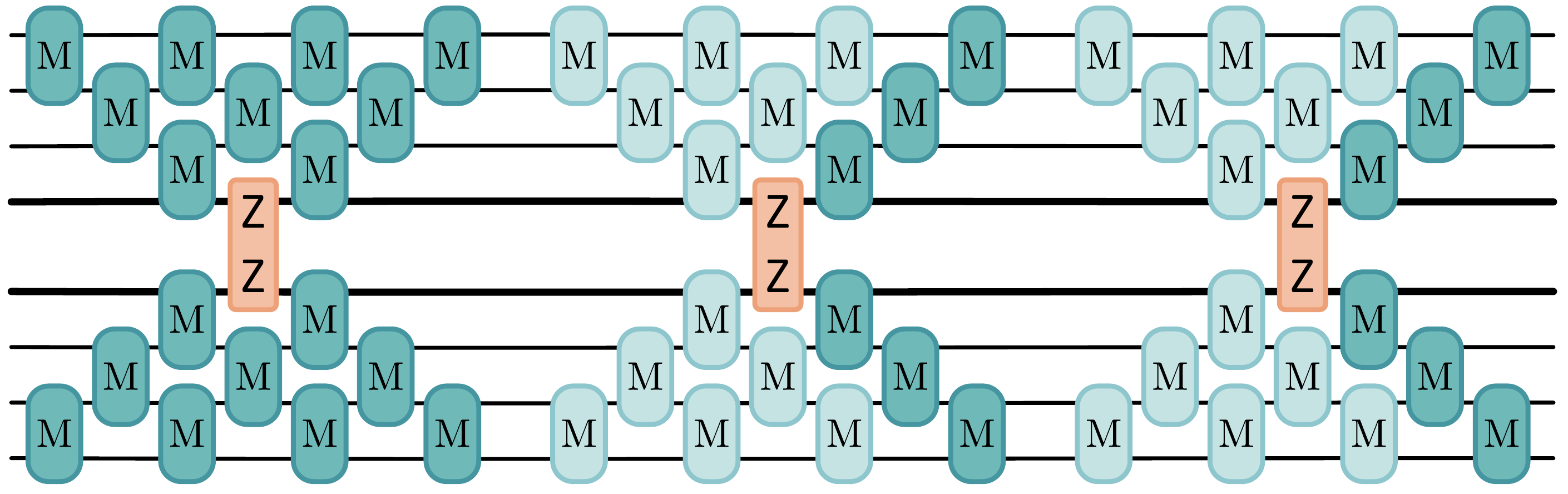}}}.
\end{align}
As can be seen, the lighter-colored blocks on the third step can be absorbed by the second step via algebraic compression. The same can be done for the second Trotter step, which leads to the following circuit  
\begin{align}\label{eq:3trottersteps_simple}
    \vcenter{\hbox{\includegraphics[height = 0.18\columnwidth]{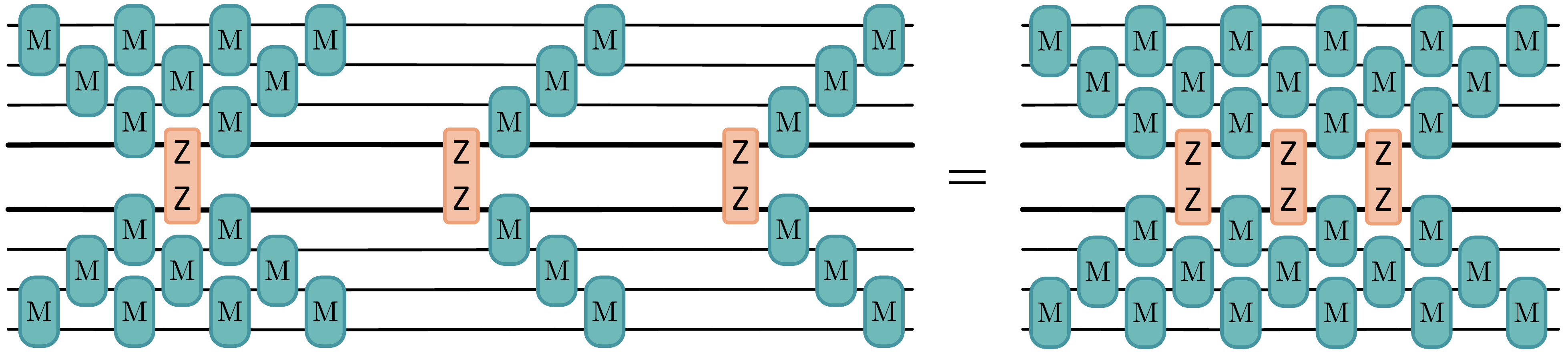}}}
\end{align}
\noindent
Consequently, in cases where the impurity is small compared to the full system size, which is commonly observed in impurity models, the cost of additional Trotter steps is greatly minimized.

\subsection{Controlled state preparation and further compression}
\label{ap: controlled state prep}

In this subsection, we will show how to build the state preparation circuit at the beginning of the time evolution circuit, i.e., the controlled $U_j$ and anti-controlled $U_j$. In this work, we are representing the ground state as a linear combination of FGS. This means that the unitaries $U_i$ can be written as free fermionic evolution operators, and controlled $U_i$ and anti-controlled $U_j$ are controlled free fermionic evolutions with a single ancilla. Using the $Q-$compression algorithm~\cite{kokcu2023algebraic}, we can compress these unitaries into a diamond structure, and obtain the following circuit:
\begin{align}\label{eq:state_prep}
    \vcenter{\hbox{\includegraphics[height = 0.20\columnwidth]{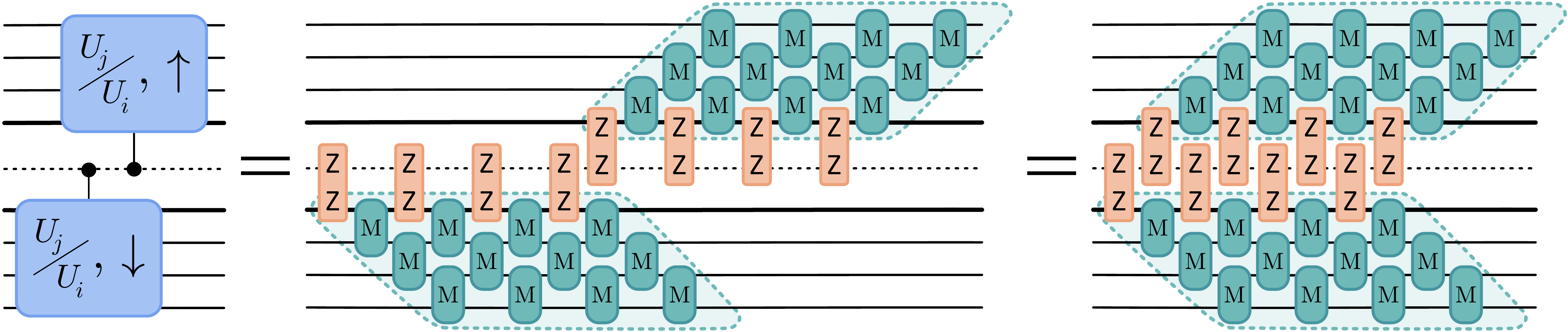}}}
\end{align}
Here, the middle qubit is the ancilla qubit (dashed line), the upper qubits are the spin $\uparrow$ qubits, and the lower qubits are the spin $\downarrow$ qubits. As the time evolution circuit contains two triangle (one reversed) structures, this circuit contains two diamond structures. Due to the conservation of number of spin $\uparrow$ and spin $\downarrow$ particles separately, we chose the FGS $\ket{\phi_i}$ as a tensor product of a FGS on each spin, i.e. $\ket{\phi_i} = \ket{\phi_i^\uparrow} \otimes \ket{\phi_i^\downarrow}$. This keeps the state preparation circuit separate and leads to two diamond structures, which is depicted in the middle of~\cref{eq:state_prep}. Because the $ZZ$ terms in each spin sector commute, we can combine the state prep into the compact form shown on the right of~\cref{eq:state_prep}.

Combining this state preparation circuit and the simplified time evolution circuit in~\cref{eq:3trottersteps_simple}, we can implement the simplified Hadamard test circuit~\cref{eq:simple_circ} which calculates the correlation $\bra{0} U_i^\dagger \hat{\gamma}_a(t) \hat{\gamma}_b  U_j  \ket{0}$ as the following:
\begin{align}\label{eq:full_circuit_raw}
    \vcenter{\hbox{\includegraphics[height = 0.20\columnwidth]{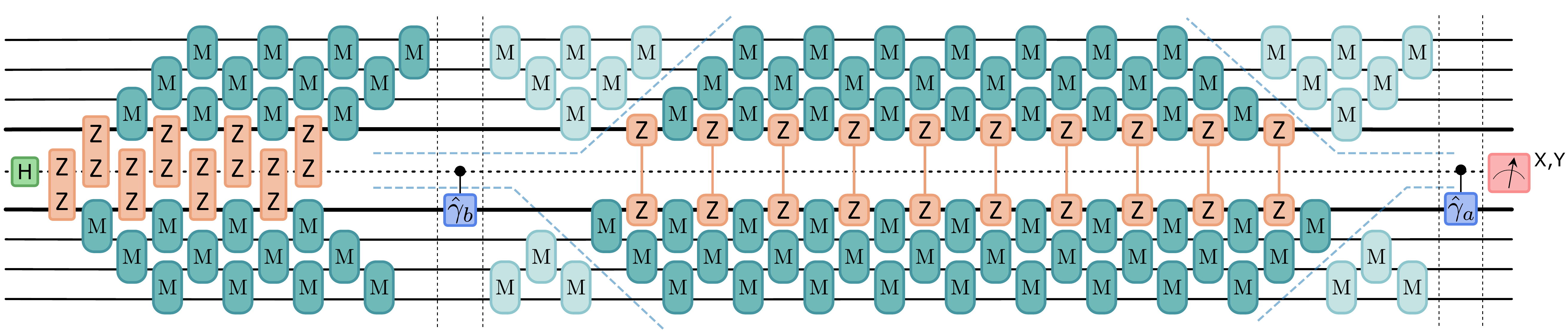}}}
\end{align}
Here we have illustrated $r=10$. Note that the impurity $ZZ$ term is split since the middle qubit is the ancilla qubit. 

As illustrated by the lighter-colored matchgates in~\cref{eq:full_circuit_raw}, this circuit can be further simplified. Firstly, those lighter-colored matchgates at the beginning of the time evolution circuit can pass the controlled $\hat{\gamma}_b$ operator, and get absorbed by the diamond structures in the state preparation circuit. Secondly, the lighter-colored matchgates at the end of the time evolution circuit can pass the controlled $\hat{\gamma}_b$ operator as well as the measurement operation. Therefore, they do not affect the measurement result and can be discarded to reduce the gate count even further. In the end, the following is the final circuit structure we ran on hardware (for $r$=10):
\begin{align}\label{eq:full_circuit_simple}
    \vcenter{\hbox{\includegraphics[height = 0.230\columnwidth]{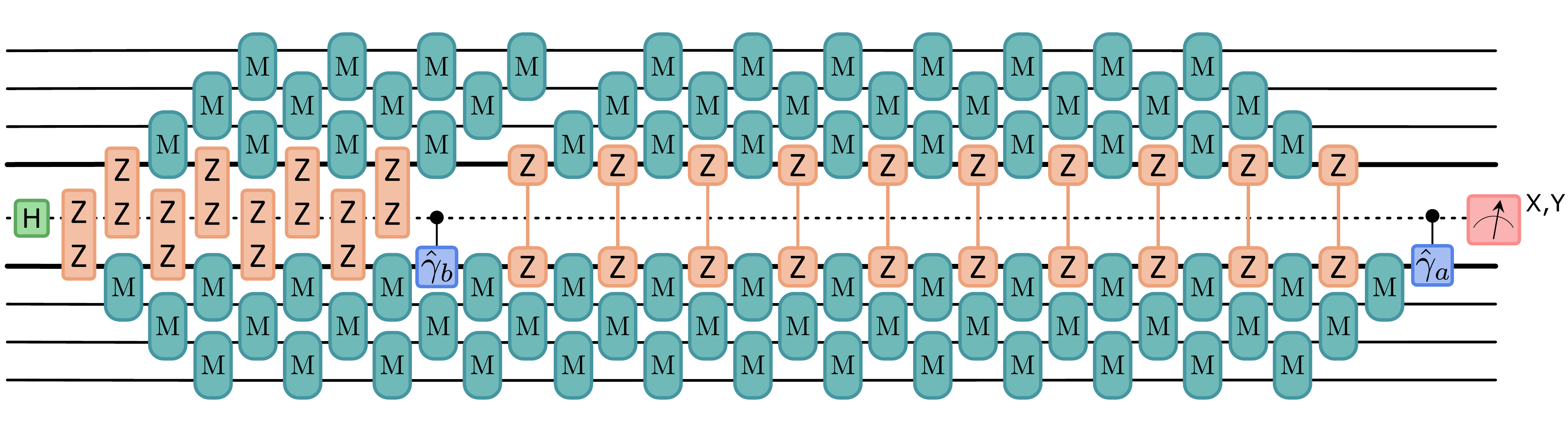}}}
\end{align}

\subsection{Quantum resource estimation}
\label{ap: quantum resource estimation}

\edit{
We now decompose the quantum resource requirements for performing DMFT into three parts: the cost per circuit, the cost per evaluation of $G^R_\text{imp}(t)$, and the cost per DMFT self-consistency cycle. The first quantifies as the number of gates per circuit, while the last two result in the number of total circuit evaluations. Each circuit evaluation's runtime (in terms of wall time) may vary depending on the hardware and number of qubits. For our hardware runs shown in~\cref{fig: sherbrooke mosaic}, the total runtime was 127~seconds for the 200 circuit evaluations ($n_t=20$ with 10 cycles of Pauli twirling).
}

\subsubsection{\edit{Per-circuit cost in terms of CNOTs}}

Using the basic structure of the circuits in this work, we compute the two-qubit gate costs per circuit as a function of the total number of bath orbitals $\fullbath$, the total number of impurity orbitals $N_I$, and the number of Trotter steps $r$. We also make use of $N_q=2(N_I+\fullbath)$, which is the number of qubits, neglecting the ancilla. The resource estimation we quote is for the most general case, which also scales the worst, where the impurity and bath qubits are fully connected. Estimation of the Trotter depths required for self-consistent solutions of the impurity solver, especially for systems beyond 30-40 qubits, is non-trivial and left for future work.

Preparing the initial states $\ket{\phi_i}, \ket{\phi_j}$ requires two (one for each spin sector) linear-depth product formula circuits as shown in~\cref{eq:state_prep}. Each exponential block, a rotation around $Z_{i}Z_{j}$ or $\alpha (X_{i}X_{j} + Y_{i}Y_{j}) + b_{i}Z_{i}+ b_{j}Z_j$, requires two CNOT gates. The state preparation then requires $N_q^{2}=4(N_I+\fullbath)^2$ two-qubit gates. The subsequent second-order Trotter steps each require an evolution of the impurity interaction and the compressed free-fermion evolution. The former includes all-to-all interactions between each orbital and spin configuration, requiring $2N_I(2N_I-1)$ CNOTs. The free-fermion evolution reduces to a ladder of matchgates with length $\fullbath$ and width $N_I$, and hopping within the impurity qubits requiring $N_I(N_I-1)/2$ matchgates. One copy of the circuit is needed for each spin sector, and each matchgate again requires two CNOTs, giving a total cost per trotter step of $6N_I^2+4N_I(\fullbath-1)$.

Next, there are an additional 2 CNOTs required to implement the controlled-Pauli operators $\hat{\gamma}_a$ and $\hat{\gamma}_b$, and an additional single layer of $\frac{N_q}{2}-1$ matchgates in a single spin sector to account for the elements which do not commute with $\hat{\gamma}_b$. Finally, there are additional circuit reductions due to the circuit light-cone of the measurement. For one spin sector, a full triangle of $\frac{N_q}{2}\left(\frac{N_q}{2}-1\right)$ CNOTs can be neglected, while in the other spin sector $\left(\frac{N_q}{2}-1\right)\left(\frac{N_q}{2}-2\right)$ CNOTs can be neglected. Thus, for circuits of Trotter depth $rN_I\geq \frac{N_q}{2}-1$, the final gate cost for circuits of the form shown in~\cref{eq:full_circuit_simple} is:
\begin{align}
\text{CNOT}(N_I,\fullbath,r) &= 4(N_I+\fullbath)^2 + r(6N_I^2+4N_I(\fullbath-1))+2 \nonumber \\
&-2(N_I+\fullbath)^2+4(N_I+\fullbath)-2 \nonumber \\
&=2N_I^2+2\fullbath^2+6N_I\fullbath+4N_I+4\fullbath \nonumber \\
&+r(6N_I^2+4N_I(\fullbath-1)).
\end{align}

The circuit depths required scale linearly in Trotter-depth and $N_IN_q$ rather than $N_q^{2}$. As $N_I \ll N_q$ in most cases of interest, this provides a quadratic reduction in gate costs.
For this work, with $N_I=1, N_B= 3$, the largest circuit used has $r=18$ and 306 CNOT gates before additional generic compiling. 

As a comparison, Ref.~\cite{jamet2022quantum} estimates that a $16$ qubit single-impurity model ($N_I= 1, \fullbath=7$) using circuits of up to 6 Trotter steps and a CNOT depth of 254, which is approximately two thousand CNOT gates. Although that work computes expectations of the form $\bra{\phi_{i}}U^{\dagger}(t_l)^{\dagger} \hat{A }U(t_{m})\ket{\phi_{j}}$, and as such is not a direct comparison, our circuit structure for a Trotter depth of 6 for this model size requires only 354 CNOTs.  In a similar study, \cite{jamet2025anderson} estimate the resources required to prepare the ground state using matrix product states compiled onto a quantum circuit. For $20$ and $40$ qubit models ($N_I= 3$ and $\fullbath=7,17$), they estimate that to prepare an initial quantum state with a fidelity of $0.99$ to the ideal ground state, one would require circuits of approximately one thousand two-qubit gates in the 20 qubit case and four thousand two-qubit gates in the 40 qubit case. Using the method presented in this work, for the same total gate counts, both state preparation and time evolution with approximately 6 and 12 Trotter layers could be computed for the 20 and 40 qubit cases, respectively. Again, the comparison is not direct, as the SGS basis used in this work requires an additional factor of $\chi^2$ circuit evaluations to produce expectations with respect to the ground state, compared to a single state preparation using matrix product state methods. Nevertheless, our CNOT count is amenable to today's noisy hardware, even for larger system sizes.

\subsubsection{\edit{Per-Green's-function cost}}

\edit{Reconstructing one retarded impurity Green's function $G^R_\text{imp}(t)$ requires $\chi^{2} n_t$ circuit evaluations, where $\chi$ is the number of fermionic Gaussian states in the SGS and $n_t$ is the number of measured time-domain samples. The $\chi^{2}$ scaling reflects that each ordered pair of FGS in the SGS contributes an independent matrix element to the decomposition of $G^R_\text{imp}$ in the SGS basis, with each matrix element requiring its own Hadamard-test circuit. Error-mitigation overhead --- Pauli twirling and zero-noise extrapolation --- multiplies this base count further: in the hardware demonstration of Fig.~\ref{fig: sherbrooke mosaic}, 10 Pauli-twirling cycles brought the total per-data-point circuit count from $\chi^{2} n_t = 20$ to 200.}

\edit{Both $\chi$ and $n_t$ can be bounded in advance for a given problem. The Bravyi-Gosset complexity result~\cite{bravyi2017complexity} bounds $\chi \sim \exp[O(b^{3})]$ at fixed impurity size and additive precision $\epsilon = 2^{-b}$, while empirical experience with selective configuration interaction methods suggests polynomial growth in the active-space dimension for impurity sizes of practical interest. The number of measured time-domain samples $n_t$ is, importantly, decoupled from the frequency resolution required of the final spectrum: the PSD method introduced in~\cref{sec: hardware runs} uses the positive-definite extension of response functions to synthesize a long signal from a short measurement, rather than the much larger Fourier-transform sample count required to resolve sharp features.}
\edit{
All $\chi^{2} n_t$ circuit evaluations contributing to a single $G^R_\text{imp}(t)$ are mutually independent and may be executed in parallel across multiple quantum processors or concurrently within a single device, making the $\chi^{2}$ factor amenable to direct wall-clock reduction on multi-QPU infrastructure.}

\subsubsection{\edit{Per-DMFT-loop cost}}
\edit{
The remaining factor in estimating total resource cost is the number of DMFT self-consistency iterations required to reach convergence, which is not known a priori for any given problem. Convergence is typically assessed by the change in the impurity model's bath parameters between successive iterations falling below a chosen tolerance. In the DMFT literature~\cite{caffarel1994exact,gull2011continuoustimemontecarlo,georges1996dynamical}, one can typically expect a few iterations to converge away from phase boundaries, with the rate set by the mixing of parameters between successive iterations~\cite{johnson1988mixing}.
}

\edit{
Near phase boundaries --- for example, the Mott transition in the half-filled single-band Hubbard model --- the convergence rate slows substantially, and tens to hundreds of iterations are not uncommon. This critical slowing down reflects the marginal stability of the DMFT self-consistency map as the system approaches the transition. Acceleration via mixing schemes can substantially reduce the iteration count in this regime, but does not eliminate the underlying slowdown.
}

\edit{
For DMFT scans across interaction strength, a ``warm start'' --- initializing each parameter point with the converged solution from a nearby point --- typically reduces the iteration count to a handful per parameter step, and to a single iteration in favorable cases when the parameter step is small relative to the local stability of the fixed-point. The total iteration count for a full phase diagram is therefore typically only modestly larger than that for a single converged solution. As a working estimate, a conservative upper bound on the per-problem iteration count for multi-orbital impurity systems near critical points is $\mathcal{O}(100)$, while warm-started scans away from criticality may require only $\mathcal{O}(10)$ or fewer.
}

\section{Quantum error mitigation and signal processing}\label{ap: post-processing}

Obtaining the hardware results presented as the magenta ``De-noised \& Extended" curve in~\cref{fig: sherbrooke mosaic}(c) involved several levels of refinement to achieve a signal from which fruitful analysis could occur. A combination of error mitigation performed on the hardware and classical post-processing was required to extract meaningful correlation functions in the frequency domain. In this section, we will explore in more detail the techniques used and the motivations behind them.

\subsection{Gate-based error mitigation}

Once the measurement of our quantum circuits occurs, the information lost to qubit dephasing during runtime cannot be fully recovered with classical post-processing techniques. To mitigate some of that dephasing, we employed dynamical decoupling (DD)~\cite{viola1999dynamical}. At the cost of a few single-qubit Pauli operations, DD acts as a deterrent to dephasing due to qubit idling. While some active qubits undergo long two-qubit operations, those that are idle will become susceptible to coupling to the environment, leading to errors. DD will act as a deterrent to these interactions by inserting a series of time dependent pulse sequences (which together amount to the identity operation) on the idle qubits. In practice, the single-qubit gate operations of a DD pulse sequence are orders of magnitude shorter than two-qubit gate operations, resulting in the benefits of including DD greatly outweighing the single-qubit gate costs.

\begin{figure}[h]
    \centering
    \includegraphics[width=0.6\linewidth]{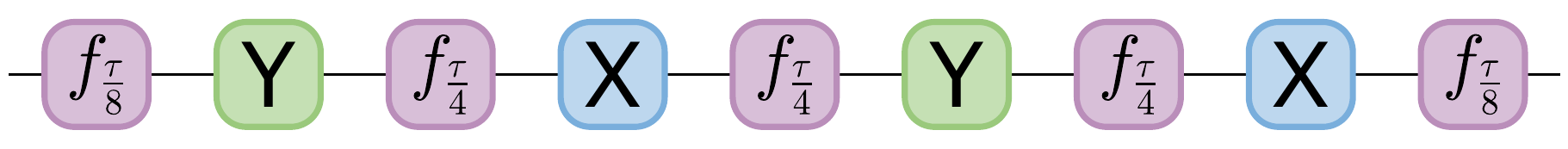}
    \caption{\textbf{$\mbf{XY4}$ pulse sequence.} The purple $f$ ``gates" represent the delay periods between the $X$ and $Y$ Pauli gates, with $\tau$ being the total idle time.}
    \label{fig: DD pulse}
\end{figure}

For our hardware runs, we employed the $XY4$ pulse sequence described by~\cref{fig: DD pulse}. Here, $\tau$ is the total idle time of the qubit, and $X,Y$ are single-qubit Pauli operations. In the time period for which we performed our hardware runs, IBM's runtime service gives access to time dependent information in transpiled circuits, which scans for idle periods. We took advantage of the Qiskit Research~\cite{qiskit2023research} Python package to implement the DD sequence.

Although qubit dephasing can be alleviated by the inclusion of DD, coherent errors due to the bias of the noisy quantum hardware are still prevalent. During the execution of a quantum circuit, if a specific gate introduces an error, say from miscalibration, repeatedly executing this gate results in compounding (or coherent) errors. To counteract coherent errors, we used an error mitigation technique called Pauli twirling, sometimes referred to as randomized compiling~\cite{hashim2021randomized, wallman2016noise}.

Pauli twirling is similar to DD, such that it introduces a few (relatively short) single-qubit Pauli operations into our circuits. A randomly selected set of Pauli gates sandwiches all two-qubit gates as a means to locally transform the state of the qubits before the two-qubit gate operation, yet keeps the logical operation of the two-qubit gate unchanged. Several instances of the same circuit are generated, each with different randomly implemented Pauli-twirled two-qubit gates, and the resulting circuits are averaged over. This is essentially changing the coherent errors in our circuit due to the repeated use of low-fidelity two-qubit gate operations into statistical errors. For our hardware results, we averaged over 10 Pauli-twirled circuits per data point in~\cref{fig: sherbrooke mosaic}. 

\subsection{Post-selection and rescaling}

Conveniently, the impurity Hamiltonian $\widehat{\mbf{H}}_\text{imp}$ is particle-conserving, which allows us to filter out the measurements received in error in a procedure called post-selection. For our hardware runs in~\cref{fig: sherbrooke mosaic}, we used a half-filled impurity model with $N_I=1$ and $N_B=3$, meaning any shots that did not obey this particle content were certainly produced in error, and could be disregarded. Each data point corresponds to 40,000 shots (4000 shots $\times$ 10 twirls per data point), and the highest shot acceptance ratio per data point was 99.1\% (first data point), with the lowest being 68.3\% ($18^\text{th}$ data point).

These post-selected results are further improved by rescaling the data based on 1- and 2-qubit gate error rates, $\epsilon_\text{1q}$ and $\epsilon_\text{2q}$ respectively, of \texttt{ibm\_sherbrooke} at the time of our hardware runs as reported by IBM's quantum compute resources~\cite{ibmquantumresources}. The rescaling formula is given by~\cref{aeq: observable rescaling}. Here, $N_\text{1q}$ and $N_\text{2q}$ are the numbers of one- and two-qubit gates in a circuit, respectively. $\hat{O}_\text{noisy}$ in our case is the raw measurement of the correlator at a particular time, and $\hat{O}_\text{rescaled}$ is the mitigated result using rescaling.

\begin{align}\label{aeq: observable rescaling}
    \hat{O}_\text{rescaled}= \frac{\hat{O}_\text{noisy}}{(1-\epsilon_\text{2q})^{N_\text{2q}}(1-\epsilon_\text{1q})^{N_\text{1q}}}
\end{align}

\begin{figure}[h]
    \centering
    \includegraphics[width=\linewidth]{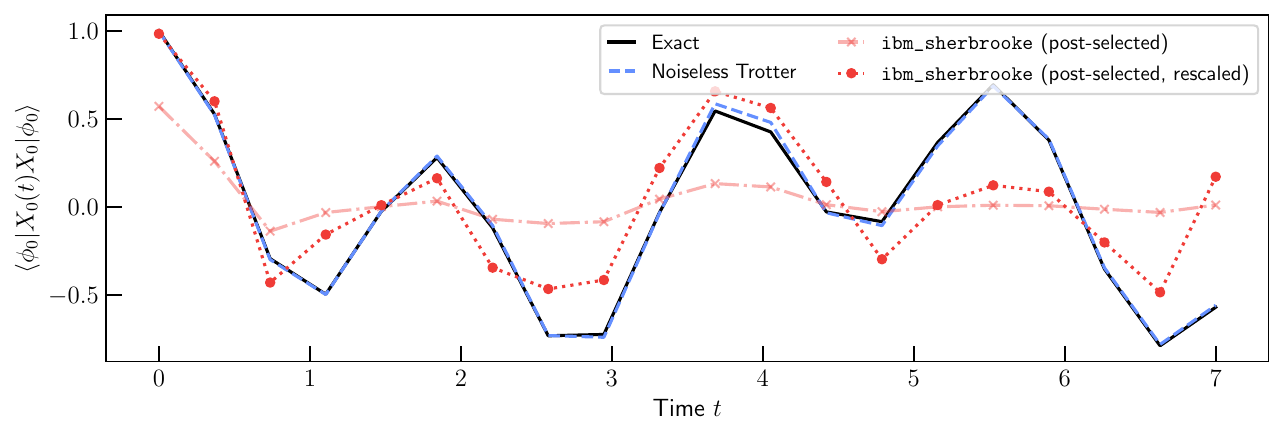}
    \caption{\textbf{Effect of rescaling based on gate error rates.} While post-selection discards known errors in our data, the amplitude of the signal is quite damped, even at early times (dashed red line, crosses). Using rescaling, much of this dampening can be mitigated (dotted red line, circles). All data is compared to the evaluation of the correlator using ED (black line).}
    \label{fig: effect of rescaling}
\end{figure}

It is important to note that all error rates reported by IBM are estimates based on randomized benchmarking~\cite{mckay2023benchmarking}, and are thus not a perfect representation of the true errors of our circuits. Nevertheless, we can consider these estimates to be the best possible without much effort in completely characterizing the errors within the circuits we ran. Further, as shown by~\cref{fig: effect of rescaling}, rescaling has a noticeable impact on recovering much of the signal that is lost due to depolarizing noise within the deeper circuits we ran on hardware. 

\clearpage

\section{\edit{Data tables and additional results}}\label{ap: tables}
\edit{As discussed in~\cref{ap: quantum resource estimation}, the number of circuit evaluations scales quadratically with the rank $\chi$ of the SGS. Thus, the practicality of a quantum simulation using the SGS crucially depends on $\chi$ remaining manageable for multi-orbital and large-bath impurity models, where quantum implementations offer the greatest advantage. Here, we concretely demonstrate that $\chi$ remains small across the system sizes considered in this work.}

\edit{Supplementing the analysis in~\cref{fig: SGS results}(a),~\cref{tab: sgs rank for fig 3a} reports the range of $\chi$ over 10 randomly sampled impurity model parameterizations to achieve the relative energy error of $\leq10^{-2}$. As described in the main text, we use particle-hole symmetric parameterizations, with baths uniquely sampled for each impurity in the multi-orbital setting. These data are also randomly sampled over interaction strengths $U$ ranging from $0$ to $9$.}

\edit{Overall, the SGS offers a greatly compacted representation of the impurity model's ground state where it matters: as the dimension of the particle-selected Hilbert space reaches 400, the fraction of the space required to achieve energy errors $\leq 10^{-2}$ drops below one percent. This advantage extends naturally to the multi-orbital setting.}

\begin{table}[h]
    \centering
    \begin{tabular}{c|c||c|c}
         $N_I$ & $N_B$ per $N_I$ & $\text{dim}(\mathcal{H}^{ps}_N)$ & $\chi$ (min--max) \\ \hline
         1 & 1 & 4 & 2--3 \\
         1 & 2 & 9 & 1--2 \\
         1 & 3 & 36 & 1--6 \\
         1 & 4 & 100 & 1--3 \\
         1 & 5 & 400 & 1--6 \\
         1 & 6 & 1225 & 1--5 \\
         1 & 7 & 4900 & 1--6 \\
         1 & 8 & 15876 & 1--4 \\
         1 & 9 & 63504 & 1--4 \\ \hline
         2 & 1 & 36 & 2--4 \\
         2 & 2 & 400 & 1--4 \\
         2 & 3 & 4900 & 1--7 \\
         2 & 4 & 63504 & 1--4 \\ \hline
         3 & 1 & 400 & 3--12 \\
         3 & 2 & 15876 & 1--8
    \end{tabular}
    \caption{\edit{\textbf{Dimension of the particle-selected Hilbert space $\boldsymbol{\text{dim}(\mathcal{H}^{ps}_N)}$ and rank window $\boldsymbol{\chi}$ of the SGS for the system sizes in~\cref{fig: SGS results}(a).} For the randomly sampled bath parameterization and interaction strengths, the required $\chi$ to achieve a relative energy error $\leq10^{-2}$ becomes an increasingly small fraction of $\text{dim}(\mathcal{H}^{ps}_N)$ as the system size grows.}}
    \label{tab: sgs rank for fig 3a}
\end{table}

\begin{figure}
    \centering
    \includegraphics[width=0.95\linewidth]{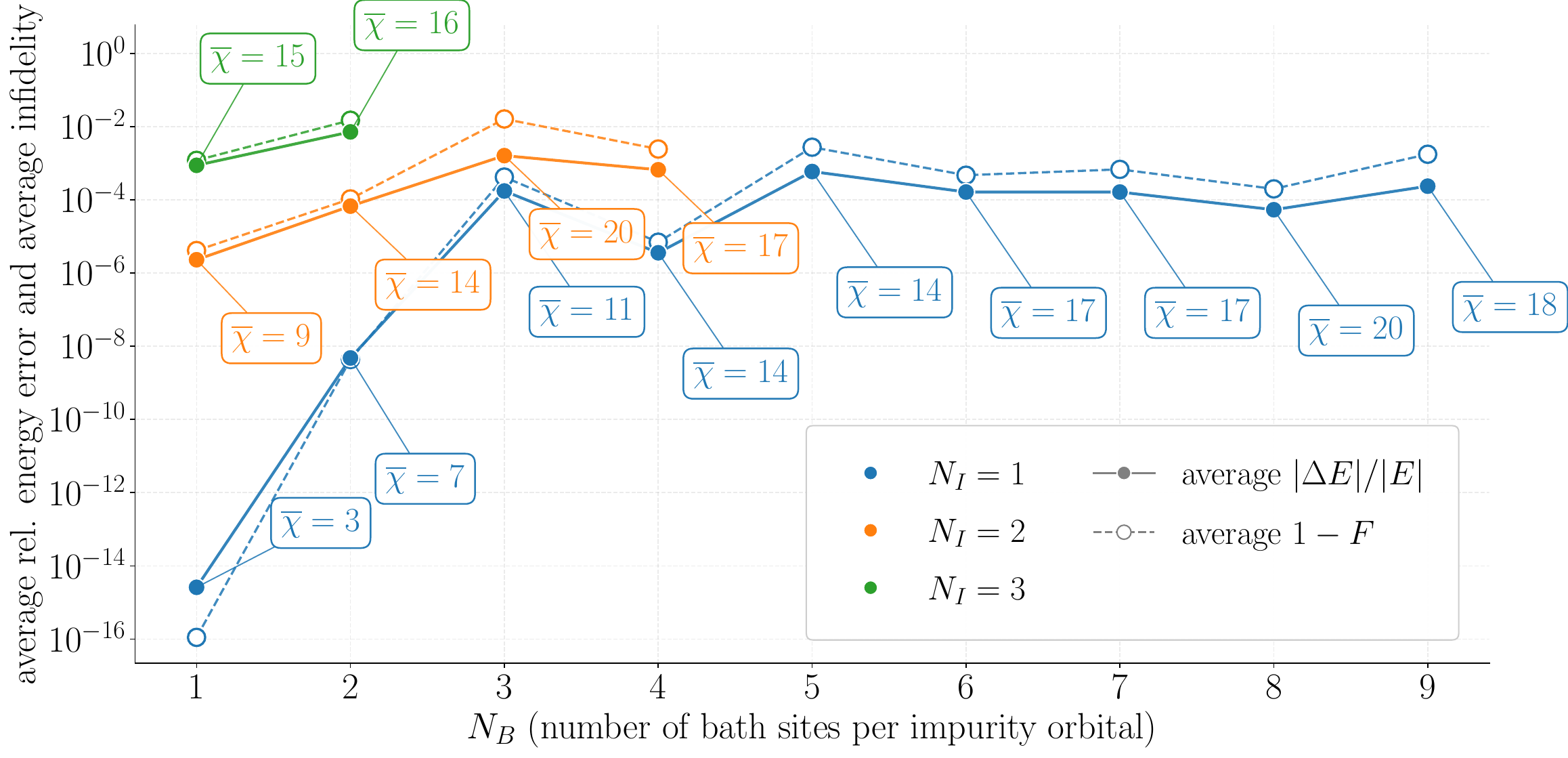}
    \caption{\edit{\textbf{Average rank and errors of the converged SGS for the system sizes considered in~\cref{fig: SGS results}(a).} Rather than fixing an energy error bound to determine the rank, the subspace search runs to convergence: adding another state to the SGS yields no further reduction in ground state energy. Averaged over 10 randomly sampled impurity models parameterized by ($U$, $V^i_b$, $\epsilon^i_b$), the average relative ground state energy error (solid line) and infidelity (dashed line) remain low despite the relatively small average SGS rank ($\overline{\chi}$ annotations) across all system sizes considered.}}
    \label{fig: more sgs scaling}
\end{figure}

\edit{Expanding on this demonstration,~\cref{fig: more sgs scaling} runs the subspace search until the termination criteria: adding another Gaussian state to the SGS produces a negligible change in ground state energy.~\cref{fig: more sgs scaling} shows the resulting average relative energy error and average infidelity over the 10 sampled models. The infidelity is computed by expanding the SGS as a state vector within the ground state's particle sector. This demonstration shows that the infidelity tracks closely with the energy error upon convergence.}

\edit{Accompanying the results in~\cref{fig: lattice GFs and QP},~\cref{tab:sgs rank for fig 4} reports the SGS rank for the DMFT-converged impurity ground state across our 1-band Hubbard model calculations on a Bethe lattice. The rank reflects the structure of the phase diagram: at small, perturbative $U$, a few FGS suffice by construction, and at large $U$ the atomic limit again admits a compact representation. Consequently, $\chi$ is peaked near the Mott transition.}

\begin{table}[]
    \centering
    \begin{tabular}{c||c}
         $U$ & $\chi$ \\
         \hline
         0.010 & 4 \\
         0.898 & 17 \\
         1.786 & 17 \\
         2.673 & 20 \\
         3.561 & 23 \\
         4.449 & 24 \\
         5.337 & 21 \\
         6.224 & 16 \\
         7.112 & 14 \\
         8.000 & 13
    \end{tabular}
    \caption{\edit{\textbf{Rank of the SGS for the DMFT results in~\cref{fig: lattice GFs and QP}.} At weak interaction, few Gaussian states suffice to represent the ground state. Similarly, the rank decreases at large $U$ as the system approaches the atomic limit, which admits a compact representation. The rank peaks near $U\approx4.449$, reflecting the critical behavior of the self-consistently determined impurity model near the Mott transition.}}
    \label{tab:sgs rank for fig 4}
\end{table}

\end{document}